\documentclass[manuscript]{acmart}
\AtBeginDocument{%
  }

\setcopyright{acmlicensed}
\copyrightyear{2018}
\acmYear{2018}
\acmDOI{}
\acmConference[Conference acronym ']{Make sure to enter the correct
  conference title from your rights confirmation email}{June 03--05,
  2018}{Woodstock, NY}
\acmISBN{}

\usepackage{xac}
\usepackage{color-edits}
\usepackage{multirow}
\usepackage{tabularx}
\usepackage{multirow}
\usepackage{subcaption}   
\usepackage{graphicx}
\usepackage[export]{adjustbox}

\usepackage{tikz}
\tikzset{box/.style={rectangle, draw, rounded corners, minimum height=1cm, minimum width=2.5cm, align=center, font=\sffamily}}

\addauthor{ys}{red}

\begin{document}

\title{\projectName{}: Seamless and Interactive Leveraging of Multiple LLMs through Comparison and Composition}

\author{Yingtian Shi}
\email{yshi457@gatech.edu}
\orcid{0000-0001-8733-7041}
\affiliation{%
  \institution{Georgia Institute of Technology}
  \country{USA}
}

\author{Jinda Yang}
% \email{}
\affiliation{%
  \institution{Simon Fraser University}
  \country{Canada}
}

\author{Yuhan Wang}
% \email{}
\affiliation{%
  \institution{Wuhan university}
  \country{China}
}
\author{Yiwen Yin}
% \email{}
\affiliation{%
  \institution{Tsinghua University}
  \country{China}
}
\author{Haoyu Li}
% \email{}
\affiliation{%
  \institution{Tsinghua University}
  \country{China}
}
\author{Kunyu Gao}
% \email{}
\affiliation{%
  \institution{Tsinghua University}
  \country{China}
}
\author{Chun Yu}
% \email{}
\affiliation{%
  \institution{Tsinghua University}
  \country{China}
}

%%
%% By default, the full list of authors will be used on the page
%% headers. Often, this list is too long and will overlap
\renewcommand{\shortauthors}{Trovato et al.}
\def \projectName {{LLMartini}}
%%
%% The abstract is a summary of the work to be presented in the
\begin{abstract}
The growing diversity of large language models (LLMs) means users often need to compare and combine outputs from different models to obtain higher-quality or comprehensive responses. However, switching between separate interfaces and manually integrating outputs is inherently inefficient, leading to a high cognitive burden and fragmented workflows. To address this, we present \projectName{}, a novel interactive system that supports seamless comparison, selection, and intuitive cross-model composition tools. The system decomposes responses into semantically-aligned segments based on task-specific criteria. It automatically merges consensus content and highlights model differences through color coding, all the while preserving unique contributions. In a user study (N=18), \projectName{} significantly outperformed conventional manual methods across all measured metrics, including task completion time, cognitive load, and user satisfaction. Our work highlights the importance of human-centered design in enhancing the efficiency and creativity of multi-LLM interactions and offers practical implications for leveraging the complementary strengths of various language models.
\end{abstract}





\begin{teaserfigure}
  \includegraphics[width=\textwidth]{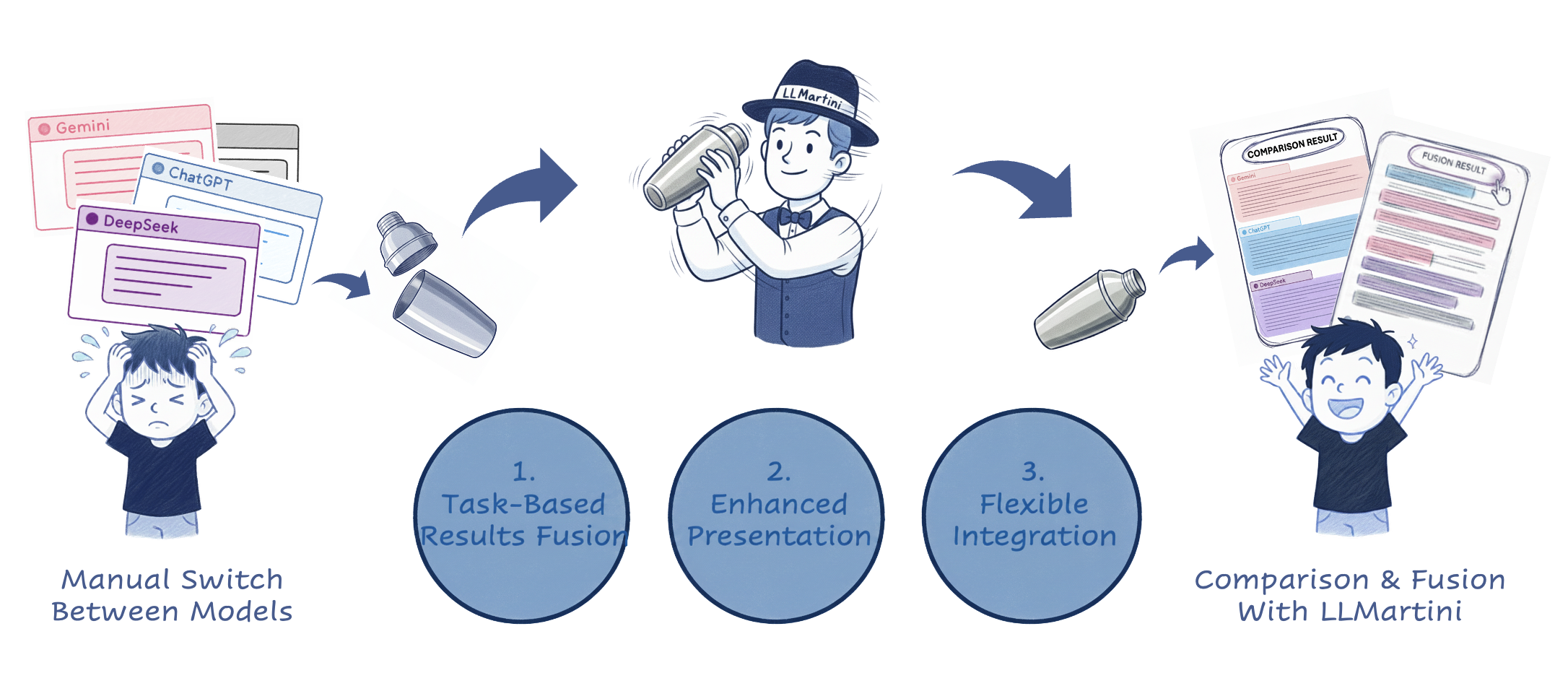}
  \caption{LLMartini is a novel interactive system that supports seamless comparison, selection, and intuitive cross-model composition. The three key features of \projectName{} are: Task-based purpose recognition and model results fusion; two enhanced interfaces for results comparison and fusion; and flexible fusion tools and functions.}
  \label{fig:teaser}
\end{teaserfigure}

\received{20 February 2007}
\received[revised]{12 March 2009}
\received[accepted]{5 June 2009}

%%
%% This command processes the author, affiliation, and title
\maketitle

\section{Introduction}
% 1. The Rise of Diverse LLMs and a New User Challenge
The rapid advancement of powerful large language models (LLMs) has fundamentally transformed how users access information and accomplish tasks \cite{bommasani2022opportunitiesrisksfoundationmodels,openai2024gpt4technicalreport}. Today, a growing diversity of models is available, each exhibiting unique strengths in areas such as factual accuracy, creative generation, or stylistic variation \cite{liang2023holisticevaluationlanguagemodels, holistic_bommasani_2023}. Faced with this expanding ecosystem, users often lack a clear understanding of each model's specific capabilities. This uncertainty has prompted users to adopt a new strategy: querying multiple LLMs with the same prompt, then comparing and combining their outputs to produce higher-quality or more comprehensive responses.
Empirical observations \cite{jakesch2023} and user studies (Section \ref{study1}) indicate that this is not a niche activity. In fact, 93\% of surveyed users actively compare outputs across different models and often engage in manual fusion, such as copying a paragraph from one model and combining it with a table from another, to construct a unified response. 

% 2. The Burden of Current Practices and the Research Gap
Despite its practical benefits, this process remains highly inefficient. It requires users to constantly switch between interfaces, mentally track sources and quality of responses, and manually integrate outputs. This fragmented process imposes a significant cognitive load, creates operational friction, and makes the integration of outputs tedious and error-prone \cite{SWELLER201137}.
Prior work has established valuable insight for Human-LLM Interaction, but remains largely confined to single-model scenarios \cite{designing_rezwana_2022, johnny_zamfirescupereira_2023, virtsi_virvou_2024}. These insights can not be directly used to support the more complex scenario of multi-LLM interaction.
Although multi-agent frameworks \cite{autogen_wu_2023, metagpt_hong_2023} employ multiple models, they typically constrain user interaction within predefined, automated workflows that prioritize system-level optimization over user agency. Similarly, LLM ensemble systems \cite{sagi2018ensemble, ensemble_mienye_2025} are designed to produce a single, optimized output through algorithmic fusion. The ensemble system often suppresses the diversity and differences that users seek when manually exploring responses across models.

% 3. Introducing Our Approach
To bridge this gap, we first conducted user studies to identify core needs and derive design principles for multi-LLM interaction. Informed by these findings, we present \projectName{}, a novel interactive system designed to support seamless and intuitive comparison, selection, and composition of outputs from multiple LLMs. 
\projectName{} replaces the fragmented multi-tab workflow with a unified interface for accessing diverse models. It introduces a dual-mode interaction paradigm: a visual comparison view for divergent responses and a fusion interface for similar ones. The system automatically segments responses into semantically aligned units, merges consensus content, highlights discrepancies via color coding, and preserves unique contributions. Through this system, users can efficiently select, combine, and refine content while significantly reducing cognitive load. 
In a user study (N=18), LLMartini significantly outperformed manual workflows across all measured metrics, including task completion time, cognitive load, and user satisfaction. Our work underscores the importance of human-centered design in facilitating efficient and creative multi-LLM interactions and offers practical insights for harnessing the complementary strengths of diverse models.

% Contributions
This work makes the following contributions:
\begin{itemize}
    \item We identify key user needs, behaviors, and pain points in multi-LLM interaction and derive actionable design guidelines for future multi-model systems.
    \item We present \projectName{}, a novel system centered on a visual comparison and user-controlled fusion paradigm, effectively reducing cognitive load and interaction friction.
    \item We conduct a user study demonstrating that \projectName{} significantly improves task efficiency, increases user satisfaction, and promotes model exploration compared to conventional multi-tab workflows.
\end{itemize}

\section{Related Work}

\subsection{Human-LLM Interaction}
With the growing capabilities of AI systems such as Large Language Models (LLMs), user interaction with AI has become more frequent and widespread. Human-LLM interaction has emerged as a critical research area in human-centered artificial intelligence. Early studies \cite{guidelines_amershi_2019, humancomputer_xu_2021, human_bingley_2022} provided foundational principles and guidelines for designing interactive AI systems, emphasizing that building effective human-AI partnerships requires enhancing transparency, controllability, and human agency. However, the uncertain capabilities and complex outputs from LLMs pose significant challenges for users \cite{reexamining_yang_2020, user_li_2023, transitioning_xu_2022}. Researchers have attempted to improve user interaction experiences through various approaches, including enhancing the interpretability and controllability of AI outputs \cite{trustworthy_li_2021,virtsi_virvou_2024,explainable_ali_2023,questioning_liao_2020,mutual_shao_2024}, developing collaborative frameworks \cite{designing_rezwana_2022}, and optimizing prompt design methods\cite{johnny_zamfirescupereira_2023, chain_wei_2022}. Extensive exploration has also been conducted on user-LLM collaboration in specific contexts \cite{investigating_guo_2024} such as healthcare \cite {humanai_sharma_2023, multimodal_lu_2024,highperformance_wang_2023}, education \cite{aigenerated_pataranutaporn_2021, human_jrvel_2023, designing_holstein_2021}, management \cite{combining_raisch_2024},  writing \cite{coauthor_lee_2022}, and programming \cite{shi2024bridginggapnaturaluser, geniewizard}.

However, the interaction paradigms proposed in these studies still focus on interaction with a single LLM. Although current research on powerful models like GPT-4 highlights their broad capabilities, it also underscores that no single model is universally suitable for all tasks \cite{sparks_bubeck_2023,exploring_reicher_2024}. Interacting with only one model limits users' ability to leverage the strengths of different models and dynamically select the most appropriate one.

To enhance the flexibility of user-LLM interaction, Luminate \cite{luminate_suh_2023} assists users in exploring diverse outputs from a single LLM. Meanwhile, emerging frameworks such as AutoGen \cite{autogen_wu_2023} and MetaGPT \cite{metagpt_hong_2023} support solving more complex tasks by composing multiple LLMs through structured dialogues, demonstrating the great potential of multi-model collaboration \cite{large_guo_2024, ai_wu_2022}.
While these multi-model frameworks provide users with more reliable outputs, they remain confined to predetermined architectures for LLM collaboration, sacrificing users' agency in comparing outputs and their control over the results of multiple models. In practical multi-LLM interactions, users tend to actively compare and combine outputs from various models—a characteristic that has not been fully explored. \projectName aims to address this gap by providing specialized interaction techniques for comparing and integrating outputs from different LLMs.

\subsection{LLM Ensemble Systems}
LLMs exhibit significant disparities in performance both on standardized benchmarks \cite{holistic_bommasani_2023,evaluating_chen_2021,bbq_parrish_2021,judging_zheng_2023} and in real-world user data \cite{crowdsourced_li_2024,wildbench_lin_2024}, reflecting an uneven distribution of capabilities across different domains and tasks. To enhance overall performance and reliability, methods that integrate multiple models have gained increasing attention. Ensemble learning has a long-established history in machine learning \cite{ensemble_dietterich_2000} and has been proven effective in improving both the comprehensive ability and output consistency of models. Correspondingly, combining multiple LLMs to achieve complementary strengths, improve answer accuracy, and enhance robustness has become an important research direction.

Various multi-model fusion strategies have been proposed in existing studies\cite{merge_lu_2024, ensemble_mienye_2025, harnessing_chen_2025}. The Mixture of Agents (MoA) framework employs a hierarchical structure that integrates generated outputs from different models—such as GPT, Claude, and Gemini—through multi-round collaborative reasoning and output fusion, significantly enhancing the accuracy and overall performance of the final output \cite{mixtureofagents_wang_2024}. LLM-Blender introduces a two-stage integration framework to improve response quality and diversity \cite{llmblender_jiang_2023}. In addition to direct output fusion methods, some studies have explored model knowledge fusion \cite{knowledge_wan_2024} and parameter fusion \cite{bridging_xu_2024, ensemble_huang_2024, uncertaintyaware_dey_2025}, both of which have demonstrated improvements in effectiveness and efficiency. Other approaches include self-integration through single-model multi-sample reasoning \cite{rethinking_li_2025}, or dynamic model selection via routing mechanisms, such as in FrugalGPT, to balance performance and cost \cite{frugalgpt_chen_2023}.

With the advancement of multi-model systems, scalable multi-LLM platforms—such as OpenAgents \cite{openagents_xie_2023} and AgentVerse \cite{agentverse_chen_2023}—have further facilitated the deployment, simulation, and evaluation of multi-model systems in real-world environments, providing support for research on model interoperability and collective intelligent behavior. Some studies suggest that well-integrated combinations of open-source models can even match or surpass the performance of proprietary models \cite{tang2025opensourcellmscollaborationbeats}.

However, existing methods still tend to provide unified outputs, often failing to adequately account for the alignment between user preferences and task characteristics. For instance, in creative tasks, users may prefer diverse candidate results rather than a single “optimal” output. Current mainstream approaches typically rely on aggregation mechanisms—such as ranking, fusion, or voting—to achieve quality optimization or consensus \cite{llmblender_jiang_2023, lv-etal-2024-urg, determinethenensemble_yao_2024}. While these strategies perform well in factual or objective tasks, they may suppress output diversity and even deviate from users’ subjective intents. Therefore, in certain scenarios, it may be necessary to intentionally maximize the diversity of generated results \cite{llmtopla_tekin_2024}.

To better adapt to user needs, we propose \projectName{}, which incorporates a more flexible model fusion mechanism—such as preference-aware routing and diversity-preserving fusion strategies—along with an optimized user interaction design. This makes the integration process more transparent and controllable. The system can dynamically adjust ensemble strategies based on explicit user feedback or implicit contextual signals, thereby delivering a more human-centric model fusion experience.

\subsection{Information Comparison and Decision Support Tools}
Decision Support Systems (DSS) have a long history of aiding users in making informed choices by presenting and analyzing relevant information \cite{past_shim_2002}. Traditional DSS often includes comparison features, allowing users to evaluate different options side by side. In recent years, machine learning (ML) and large language models (LLMs), have been widely integrated into DSS to enhance their analytical and decision-making capabilities across complex domains such as healthcare \cite{involvement_jayatilake_2021, aibased_tutun_2022,evaluating_rao_2023, aigenerated_liu_2023, leveraging_benary_2023}, business \cite{integrating_somlai_2022}, resource management \cite{design_jian_2021, chatgpt_iswahyudi_2023}, and scientific research and engineering \cite{procksi_barthel_2007, circular_yu_2022}.

Within the emerging framework of Evaluative AI, the focus has shifted from providing a single answer toward presenting evidence, enabling users to make decisions themselves through the comparison and evaluation of information \cite{explainable_miller_2023}. This shift underscores the central role of information comparison in the decision-making process. The form of DSS is being reshaped, with changes in how information is aggregated, accessed, and shared \cite{large_burton_2024, artificial_gupta_2021}.
The importance of information comparison in decision-making is well-established. Research indicates that comparing multiple pieces of information can reduce uncertainty and lead to better decisions \cite{comparing_kim_2023, value_luhede_2025}. This is particularly valuable in complex fields such as environmental protection, where single sources may be incomplete or biased. Furthermore, DSS incorporating information comparison has been widely applied in optimizing machine learning methods \cite{knowledgebased_rosati_2022} and supporting human resource decisions \cite{10.1007/s00500-021-06659-4}, among other specialized domains.
The effective comparison and utilization of multi-source information also depend significantly on its fusion and presentation. Methods for multi-source information fusion have been developed to integrate diverse information sources \cite{multisource_wu_2025, multiagent_tu_2025, FEI2026103512}, and pattern-based approaches have been proposed for context-aware knowledge fusion within DSS \cite{patterns_smirnov_2015}.

However, existing comparison and fusion techniques were not designed for the generative nature of LLM outputs. Traditional side-by-side viewers lack specialized functionalities to compare textual responses that vary in structure, style, and content. Similarly, current DSS do not address the unique challenges of comparing and integrating outputs from multiple LLMs, such as handling differing response formats, identifying complementary information, and manually consolidating content. Inspired by current DSS design principles, we developed \projectName{}, an LLM Fusion and Comparison Platform. It provides users with a systematic tool to weigh the value of different LLM outputs. Our design embodies the philosophy of Evaluative AI \cite{explainable_miller_2023} by refraining from providing a single "correct answer." Instead, based on the result analysis, it highlights differences among models, supporting users in comparison, editing, and final decision-making.
% \subsection{Research Gap}

% Our review reveals a significant gap between the technical advances in multi-model systems and the user's need to compare and combine LLM outputs. While ensemble methods offer algorithmic pathways for combining models, they operate automatically without user involvement. While evaluation benchmarks provide aggregate performance metrics, they do not assist users with task-specific decisions. While decision support systems facilitate information comparison, they are not designed for generative AI outputs.

\projectName addresses the research gap by introducing the first interactive system specifically designed to support users in comparing and fusing multiple LLM outputs. Our work integrates insights from HAI research on human control and transparency, ensemble methods for combining model outputs, decision support principles for information comparison, and evaluation methods for assessing model capabilities—creating a novel interaction paradigm that places the user at the center of multi-LLM workflows.

\section{Study1: Understanding User Behaviors and Needs in Multi-LLM Interactions}
\label{study1}

We have observed that users need to interact with multiple models simultaneously. To address this, we first conducted a questionnaire survey to quantify the scale of this demand and explore current user habits in LLM interactions. We collected 383 valid responses (159 male, 221 female, 3 prefer not to say, mean age = 23.96, SD = 3.87) from participants of diverse age groups, educational backgrounds, and professions.

\begin{figure}[htbp]
    \centering
    \begin{subfigure}{0.45\textwidth}
        \centering
        \includegraphics[width=0.95\textwidth, valign=t]{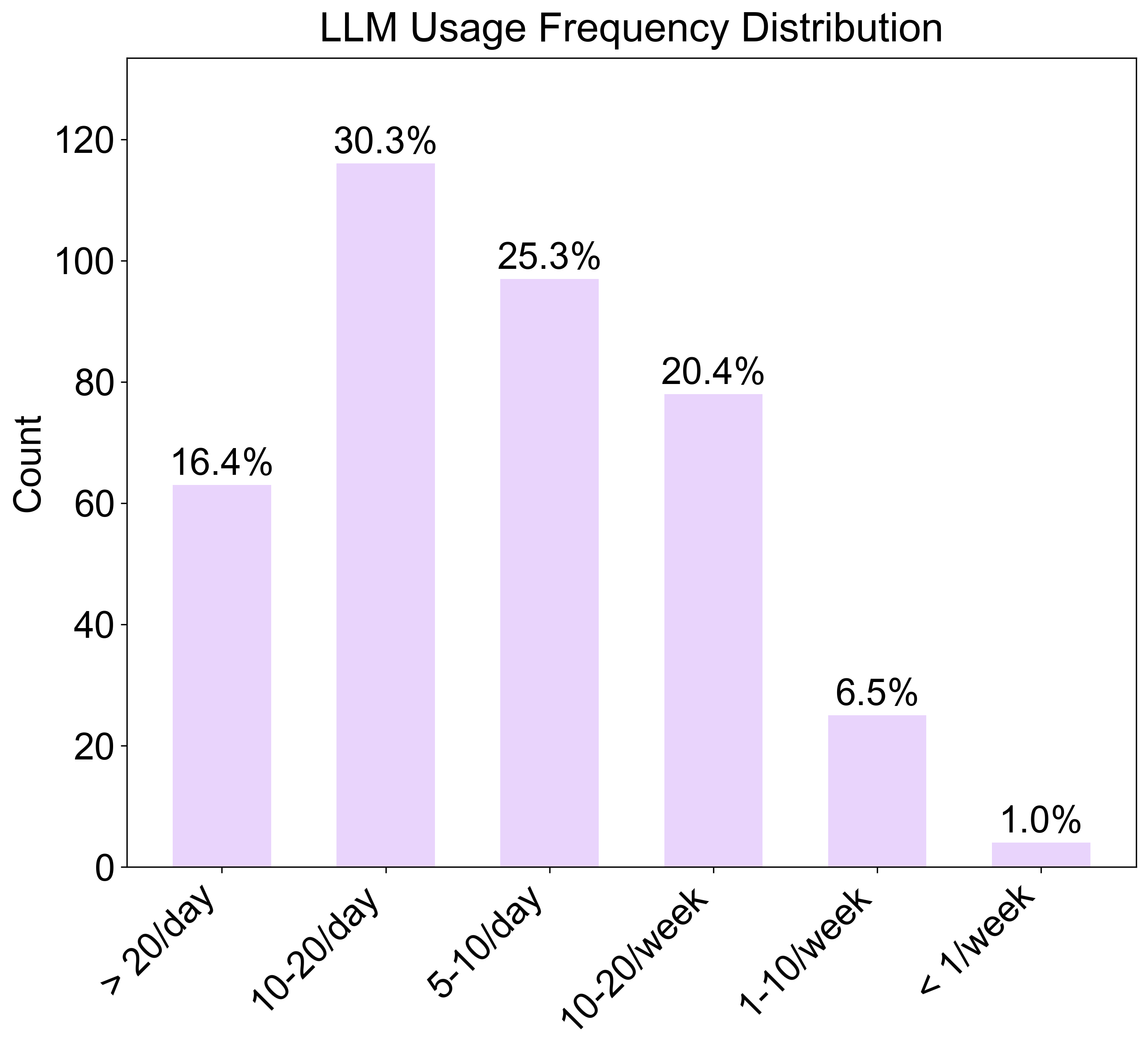} 
        \label{fig:llm_frequency_chart}
    \end{subfigure}
    \hspace{1em}
    \begin{subfigure}{0.45\textwidth}
        \centering
        \includegraphics[width=0.97\textwidth, valign=t]{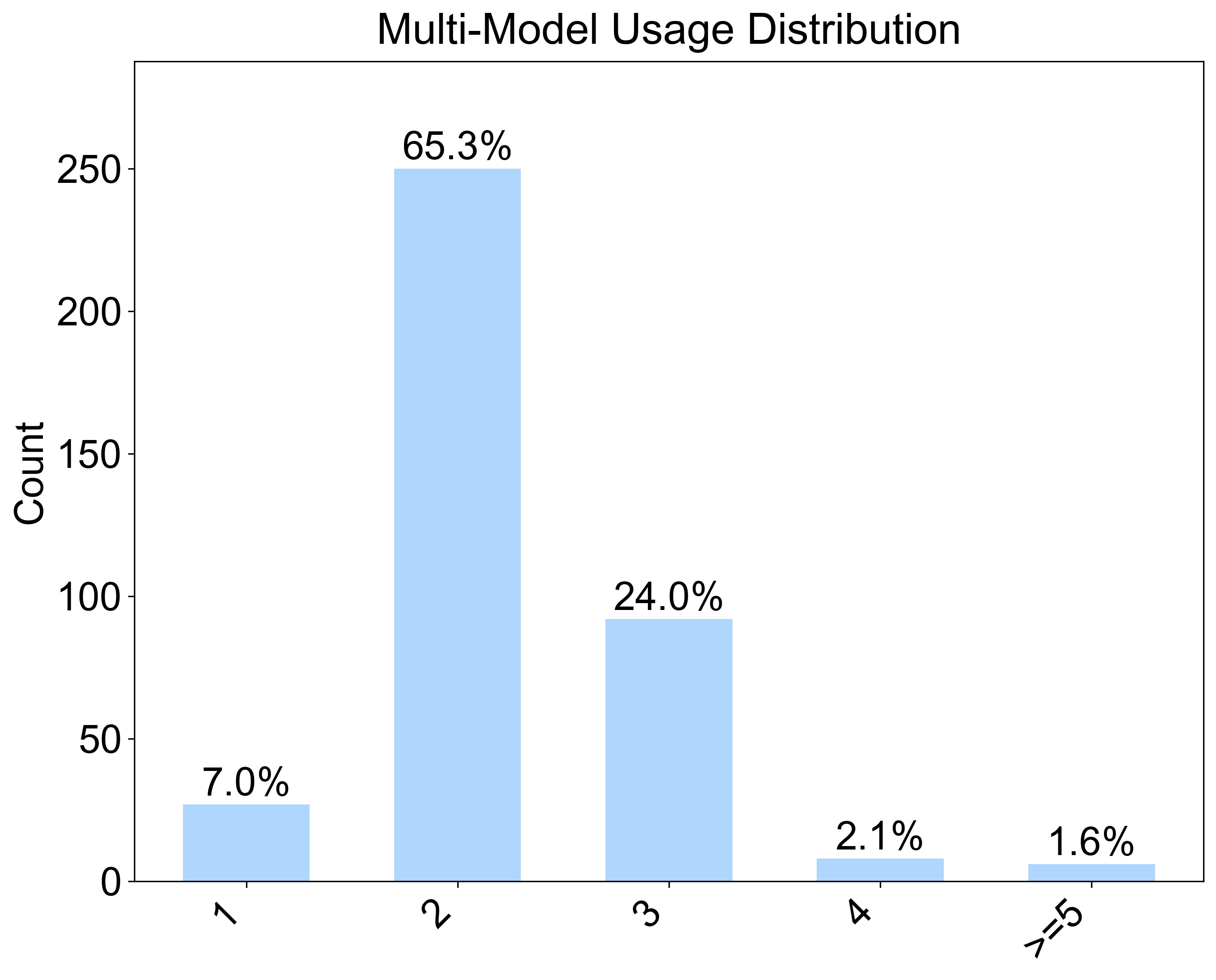} 
        \label{fig:multi_model_usage_chart}
    \end{subfigure}
    \caption{The frequency distribution of users' interactions with LLM and the number of models used simultaneously by users in daily interactions.}
    \label{fig:llm_usage_frequency}
\end{figure}

The survey results (Fig. \ref{fig:llm_usage_frequency}) reveal that 72\% of users interact with LLMs more than five times per day, with 16.4\% engaging in more than 20 interactions daily. Additionally, 93\% of users reported using two or more models concurrently, among which 64.9\% use exactly two models simultaneously. These findings highlight not only the significant demand for LLM interactions but also a strong preference for multi-model engagement.

% \yscomment{Frequency of Model Use, Number of Models Used}

To further investigate user motivations and needs in multi-model interactions, we developed a website that allows users to interact with multiple LLMs at the same time and conducted a user behavior analysis experiment. Through this experiment, we aim to address the following research questions:
\begin{itemize}
    \item RQ1: What are the core motivations for users to interact with multiple large language models?
    \item RQ2: How do task types and model characteristics influence users’ behavioral patterns when interacting with multiple models?
    \item RQ3: How do users compare, evaluate, and integrate outputs from different models? What are the main challenges and pain points encountered in this process?
\end{itemize}

\subsection{Apparatus: The Basic Multi-LLM Prototype}
We developed an experimental prototype website designed to enable users to obtain responses from multiple large language models (LLMs) simultaneously on the same page and provide feedback on the model outputs. The system uses FastAPI as the backend framework and React for the frontend interface, integrating APIs from several mainstream models. All user interaction requests and model responses are persistently stored in a local PostgreSQL database.

Based on the four most frequently used models identified in our preliminary survey—DeepSeek \cite{deepseek2024chat}, ChatGPT \cite{openai2023chatgpt}, Doubao \cite{bytedance2024doubao}, and Gemini \cite{google2024gemini}—we included them as the supported models in the system.

% \yscomment{Website Interface}
\begin{figure}[htbp]
    \centering
    \includegraphics[width=0.9\linewidth]{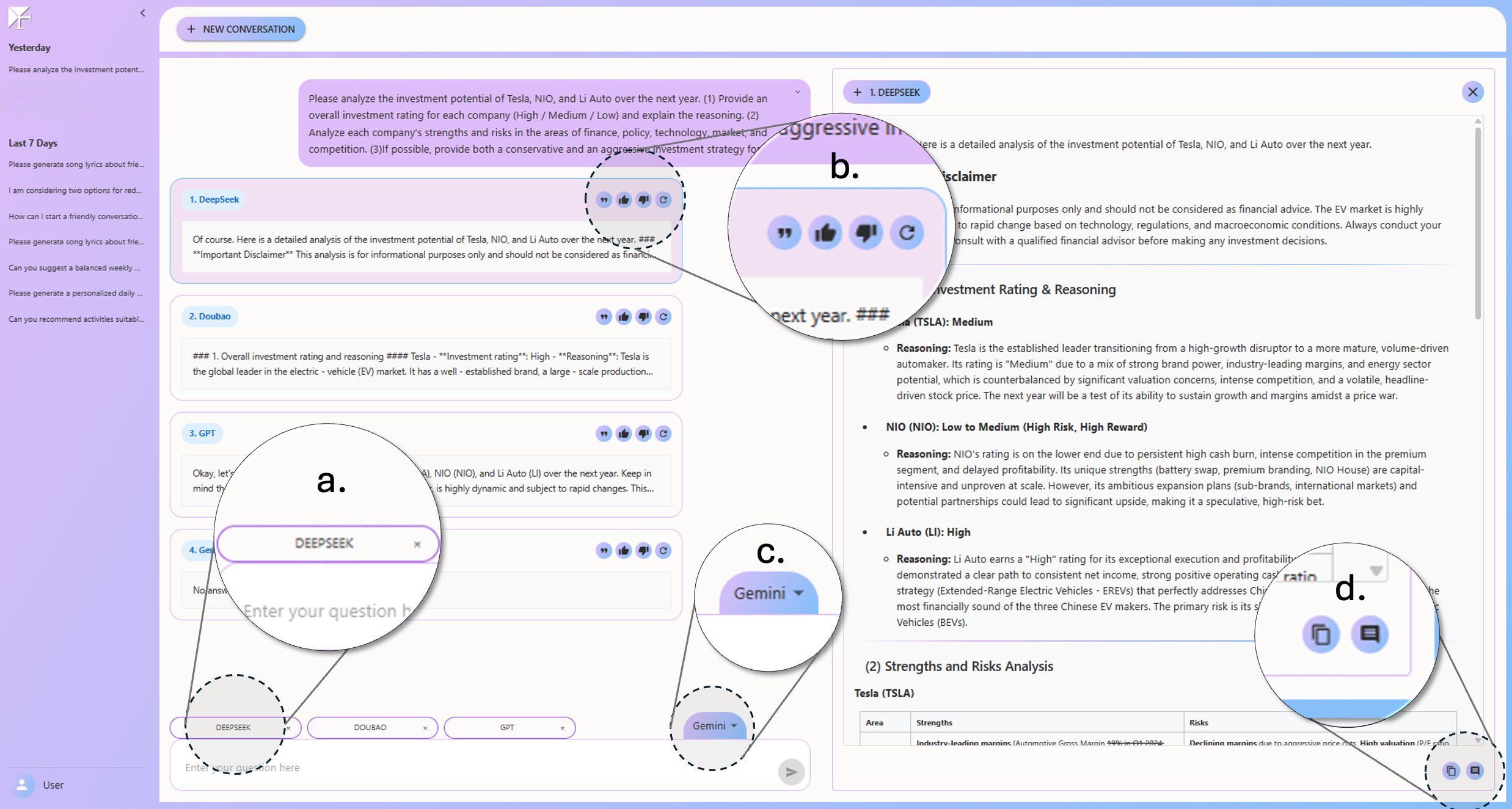}
    \caption{Website prototype, with the history bar on the left, user interaction input and multi-model cards in the middle, and model output details on the right. Zoomed highlights show: (a) cited model name, (b) four action buttons ("Cite", "Like", "Dislike", and "Regenerate") on summary card, (c) dropdown model selector, (d) "Copy" and "Comment" buttons on expanded panel.}
    \label{fig:web1}
\end{figure}

In terms of page layout, we drew inspiration from mainstream LLM products (such as DeepSeek and ChatGPT), adopting a navigation bar to display historical conversation records and a main chat interface. This design aims to closely mimic users’ familiar interaction environment and reduce the learning cost.

The output of each model is presented in two forms: a summary card and an expanded view. The summary card displays the model name and the first 50 characters of the generated content, with four action buttons at the top (Fig. \ref{fig:web1} b.): "Cite", "Like", "Dislike", and "Regenerate". Expanding the panel reveals the full response, along with support for adding comments and copying the text (Fig. \ref{fig:web1} d.).

After a user submits a query, summary cards for all models are displayed side by side. By default, the first model in the list generates a response automatically; users can also manually trigger other models to run. If unhappy with a result, they can click “Regenerate” to refresh that model’s output. All responses are streamed progressively and can be fetched simultaneously.

Users can quickly evaluate model outputs via “Like” or “Dislike” buttons, or provide more detailed textual feedback in the expanded panel. The “Cite” (Fig. \ref{fig:web1} b.) feature allows users to select one or more results as context for subsequent dialogue, and cited model names are displayed next to the input box (Fig. \ref{fig:web1} a.). If no model is explicitly selected, the response from the first model is used as the default context. Additionally, users can specify a preferred model (Fig. \ref{fig:web1} c.)before the conversation begins; its card will be automatically pinned to the top and set as the default for generating responses.

\subsection{Study Procedure}
We invited 19 users (9 male, 10 female, mean age = 25.37, SD = 4.54 ) to participate in the experiment. The participants had diverse backgrounds and varied significantly in their frequency of LLM usage and typical task scenarios. The experiment lasted for one week, during which users were asked to use our platform following their natural interaction habits. While we did not impose strict restrictions on usage scenarios, we encouraged users to cover as many different types of tasks as possible.

To ensure data quality and sufficiency, we required each user to complete at least ten distinct types of tasks and submit no fewer than 70 requests in total. Additionally, after each request, users were required to provide quick feedback (like or dislike ratings) for at least one model’s output and submit at least one written comment.

At the end of the experiment, we conducted approximately 30-minute semi-structured interviews with each user to gain in-depth insights into their experience, main pain points, and their perspectives and suggestions regarding the multi-model interaction paradigm. As a token of appreciation, each participant received a gift card worth between \$15 to \$30, based on the amount of data they contributed.

\subsection{Study Results}
We cleaned the collected data by excluding invalid records caused by system errors or model invocation failures, ultimately obtaining 1893 valid user queries and 8058 model interaction records (each successful retrieval of a model output was counted as one interaction, including regeneration). For user feedback, we filtered out short comments containing fewer than 10 characters, retaining 2049 instances of quick feedback (likes/dislikes) and 1856 valid textual comments. To further analyze user behavior and model performance in depth, we conducted a detailed examination of user queries and their corresponding model interaction records.

\subsubsection{Task-Related Analysis}
In terms of task segmentation, we distinguished task types based on user requests and their contextual information: significantly different topics within the same session were treated as independent tasks, while follow-up questions on the same topic were grouped under the same task.

Task categorization employed a heuristic iterative method: First, we selected 50 diverse tasks as seed samples for manual pre-classification. Subsequently, this classification result was used as a prompt to iteratively categorize the remaining tasks using the DeepSeek model. The model determined whether a new task belonged to an existing category; if not, it generated a new category and dynamically updated the prompt. After multiple iterations, we manually verified and adjusted the classification results to enhance inter-category distinction. Finally, all tasks were reclassified based on the stabilized prompt from the iterative process to ensure consistency. We established a classification system along two distinct dimensions: based on task content (the specific topics involved), resulting in 66 categories; and based on task purpose (the user's core intent behind the query), resulting in 4 broad categories and 17 subcategories (One of the categories is "Others"). The purpose-based classification system includes the following four typical types (each broad category contains several subcategories):
\begin{itemize}
    \item Content Generation(G): Refers to creating new text, code, or creative ideas from scratch, emphasizing originality and creativity. Includes "Creative Writing", "Opinion Expression", "Simulated Dialogue", "Code Generation", "Explanation", and "Question Generation".
    \item Content Editing(E): Involves modifying, polishing, reorganizing, or optimizing existing text to improve its readability, accuracy, and expressiveness. Includes "Text Editing", "Copywriting Editing", and "Emotional Editing".
    \item Information Retrieval (R): Aims to retrieve, filter, and integrate information from various sources based on user queries to fulfill factual inquiries, comparisons, or extraction needs. Includes "Information Query", "Comparative Analysis", and "Information Extraction".
    \item Problem Solving (S): Provides practical solutions or operational guidance through analysis, reasoning, and suggestions tailored to specific user needs. Includes "Recommendation", "Q\&A Analysis", "Exercise Solving", and "Solution Design".
\end{itemize}

% \yscomment{Content Category Distribution, Content Task Length Figure, Purpose Category Distribution, Purpose Task Length Figure}
\begin{figure}[htbp]
    \centering
    \begin{subfigure}{0.50\textwidth}
        \centering
        \includegraphics[width=0.95\textwidth, valign=t]{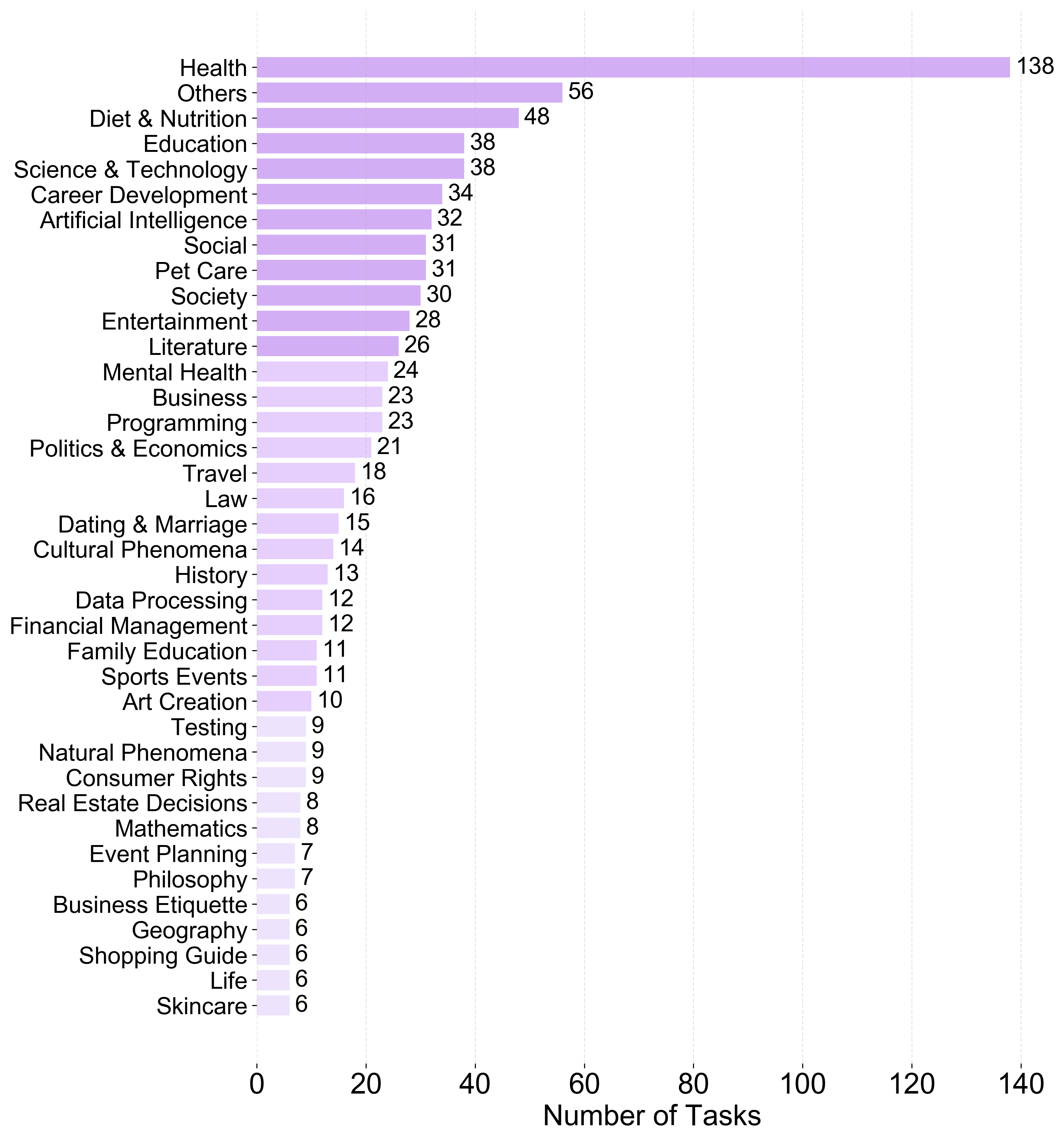} 
   
    \end{subfigure}
    \hspace{1em}
    \begin{subfigure}{0.45\textwidth}
        \centering
        \includegraphics[width=0.95\textwidth, valign=t]{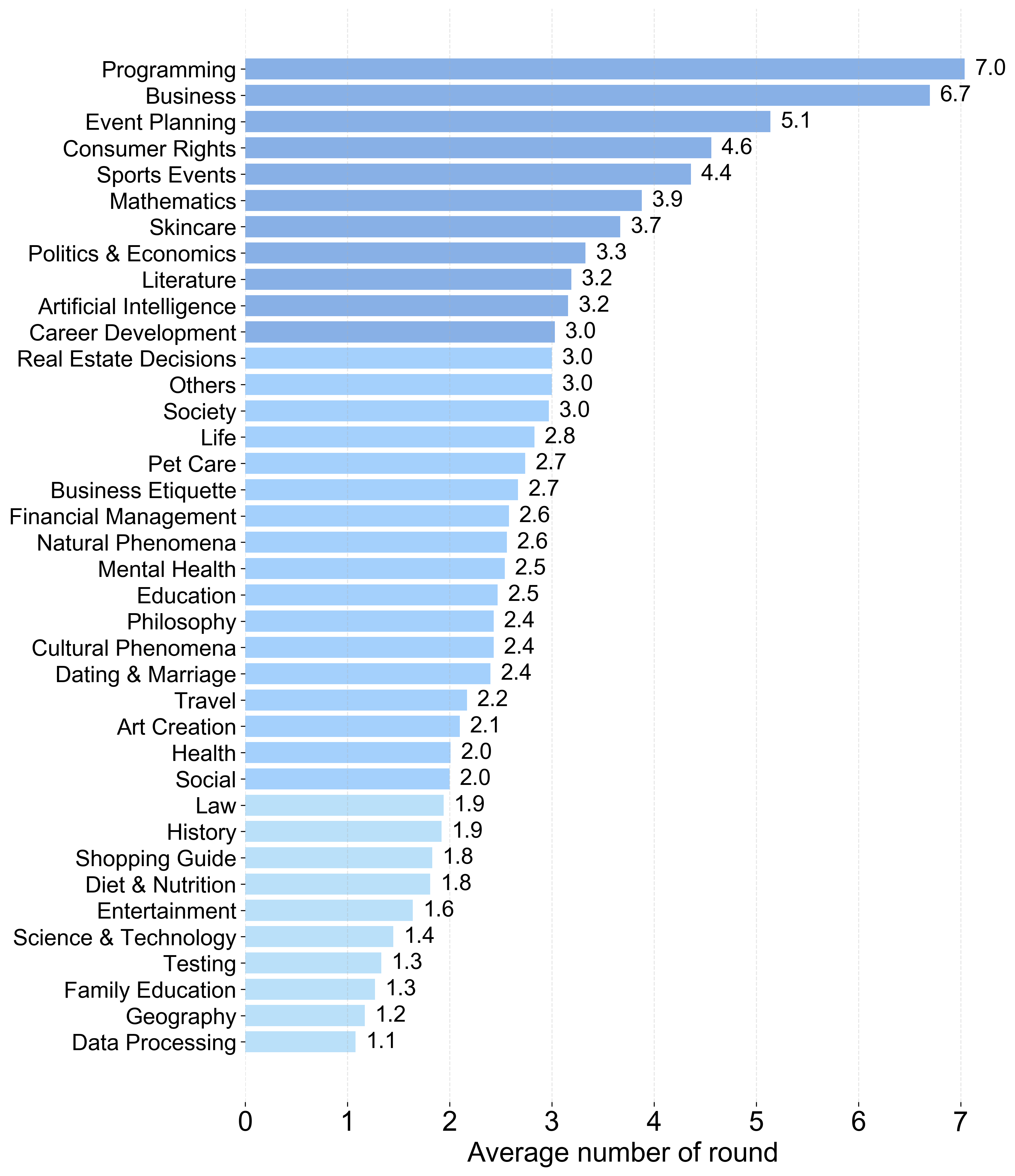} 
      
    \end{subfigure}
    \caption{Based on the statistical results of the task topic, the left figure shows the average total number of tasks for each topic, and the right figure shows the average number of user request rounds required to complete a task for the topic.}
    \label{fig:topic_statistic}
\end{figure}

The results in Figure \ref{fig:topic_statistic} indicate that the topics of the tasks are widely distributed across multiple domains. Among them, health and diet-related questions were the most frequent topics of concern for general users. In terms of interaction rounds (number of user queries per task), complex tasks such as programming, business, and event planning significantly exceeded other types, demonstrating users' deeper interactive demands in these scenarios.

\begin{figure}[htbp]
    \centering
    \begin{subfigure}{0.45\textwidth}
        \centering
        \includegraphics[width=0.95\textwidth, valign=t]{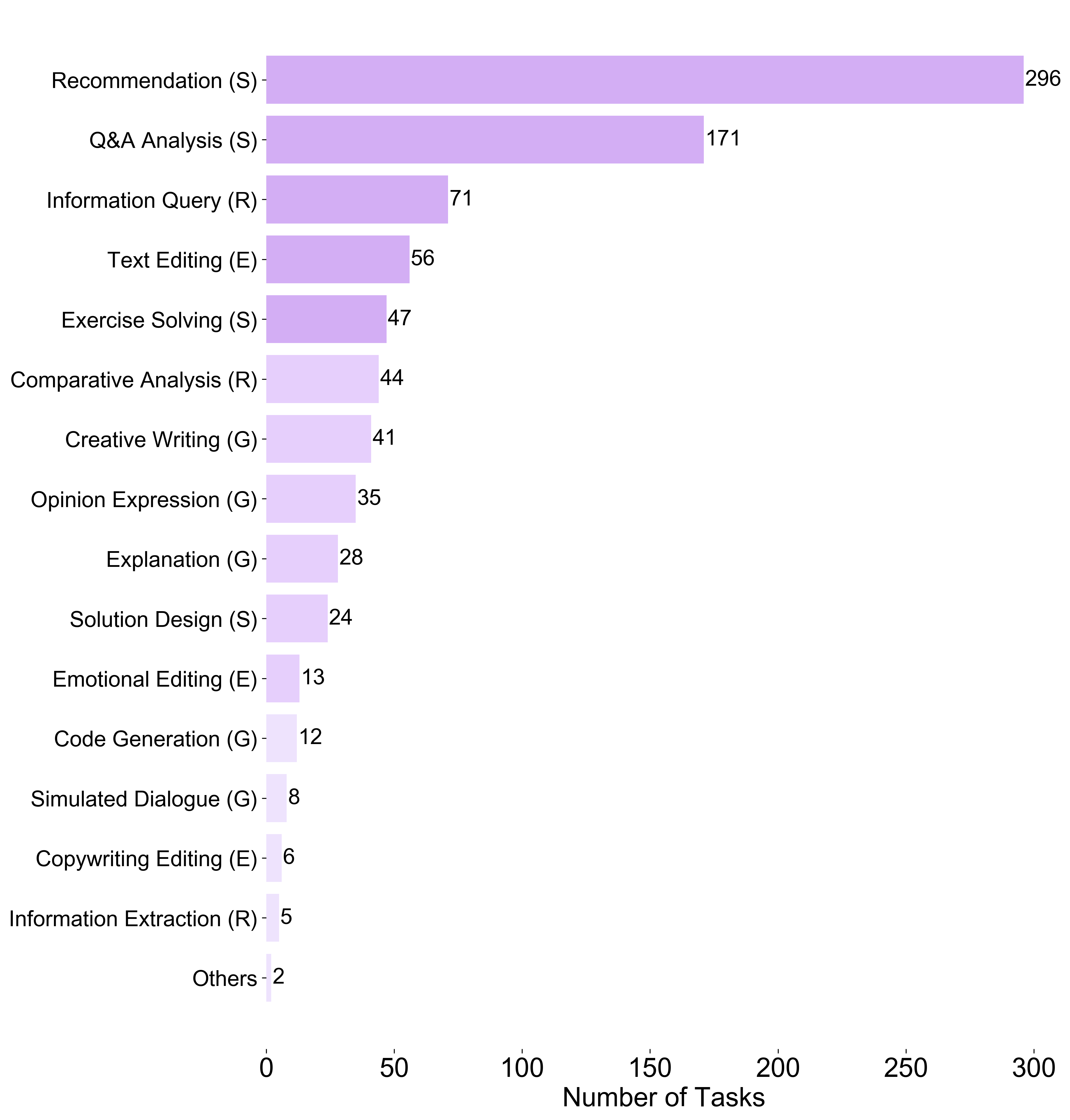} 
   
    \end{subfigure}
    \hspace{1em}
    \begin{subfigure}{0.45\textwidth}
        \centering
        \includegraphics[width=0.95\textwidth, valign=t]{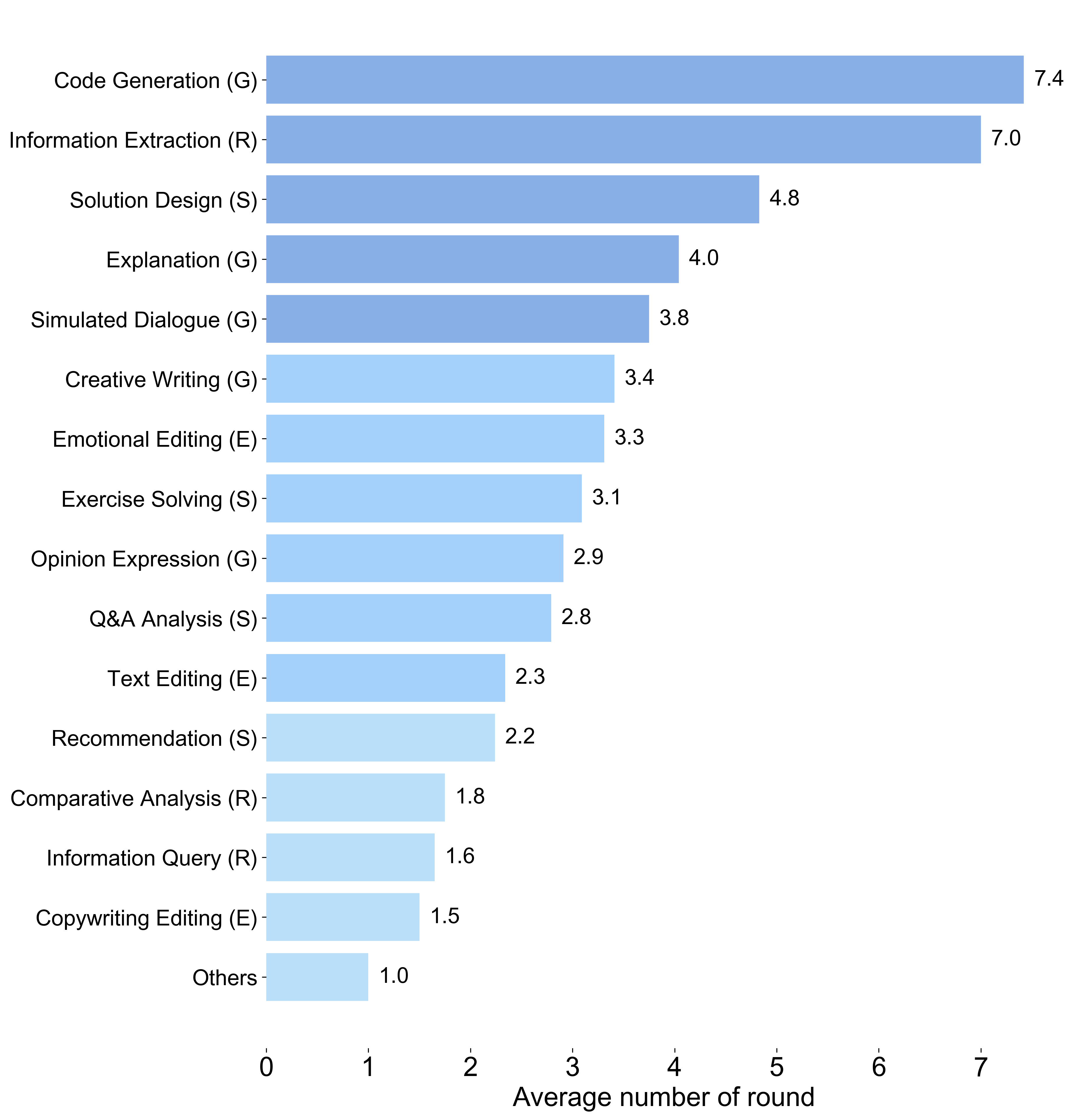} 
      
    \end{subfigure}
    \caption{The statistical results based on the task purpose are as follows: the left side shows the average number of tasks for each topic, and the right side shows the average number of user request rounds required to complete a task for that topic. The capital letters after the task classification represent the major categories to which it belongs.}
    \label{fig:type_statistic}
\end{figure}

From the distribution (Fig. \ref{fig:type_statistic}), Problem Solving tasks accounted for 61.91\%, making them the primary purpose of user interaction, indicating that users tend to use large language models as tools for addressing practical problems. However, Content Generation tasks had a relatively lower proportion; their average number of interaction turns was significantly higher, suggesting that these tasks often require multiple rounds of iteration and refinement.

\subsubsection{Model-Related Analysis}
\begin{figure}[htbp]
    \centering
    \includegraphics[width=0.8\linewidth]{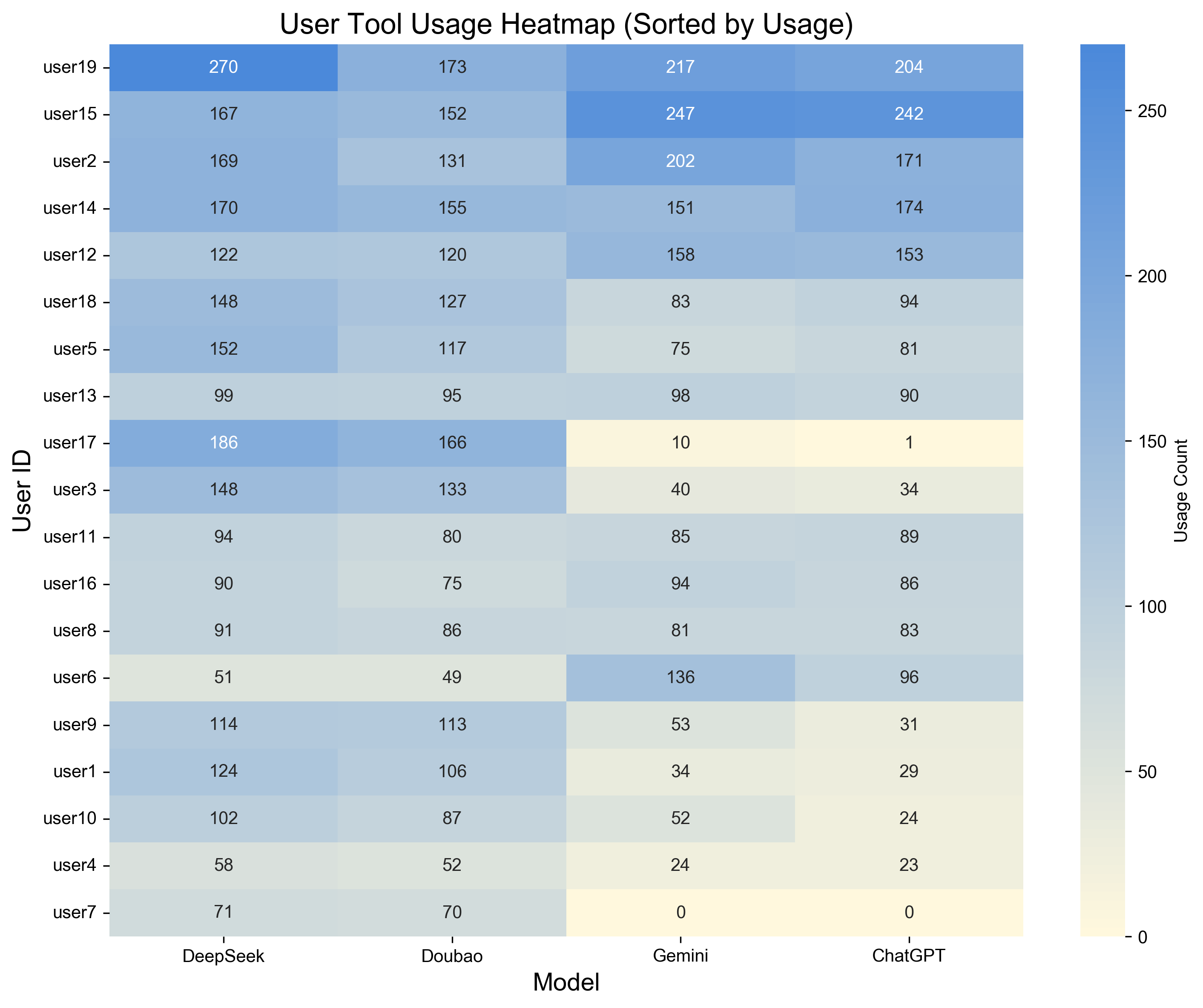}
    \caption{The heat map of model interaction (Darker blue indicates more usage, Lighter yellow indicates less usage). The results are sorted from top to bottom according to the total number of user interactions. The number in each block represents the number of interaction times.}
    \label{fig:heapmap}
\end{figure}
\begin{figure}[htbp]
    \centering
    \begin{subfigure}{0.8\textwidth}
        \centering
        \includegraphics[width=\textwidth, valign=t]{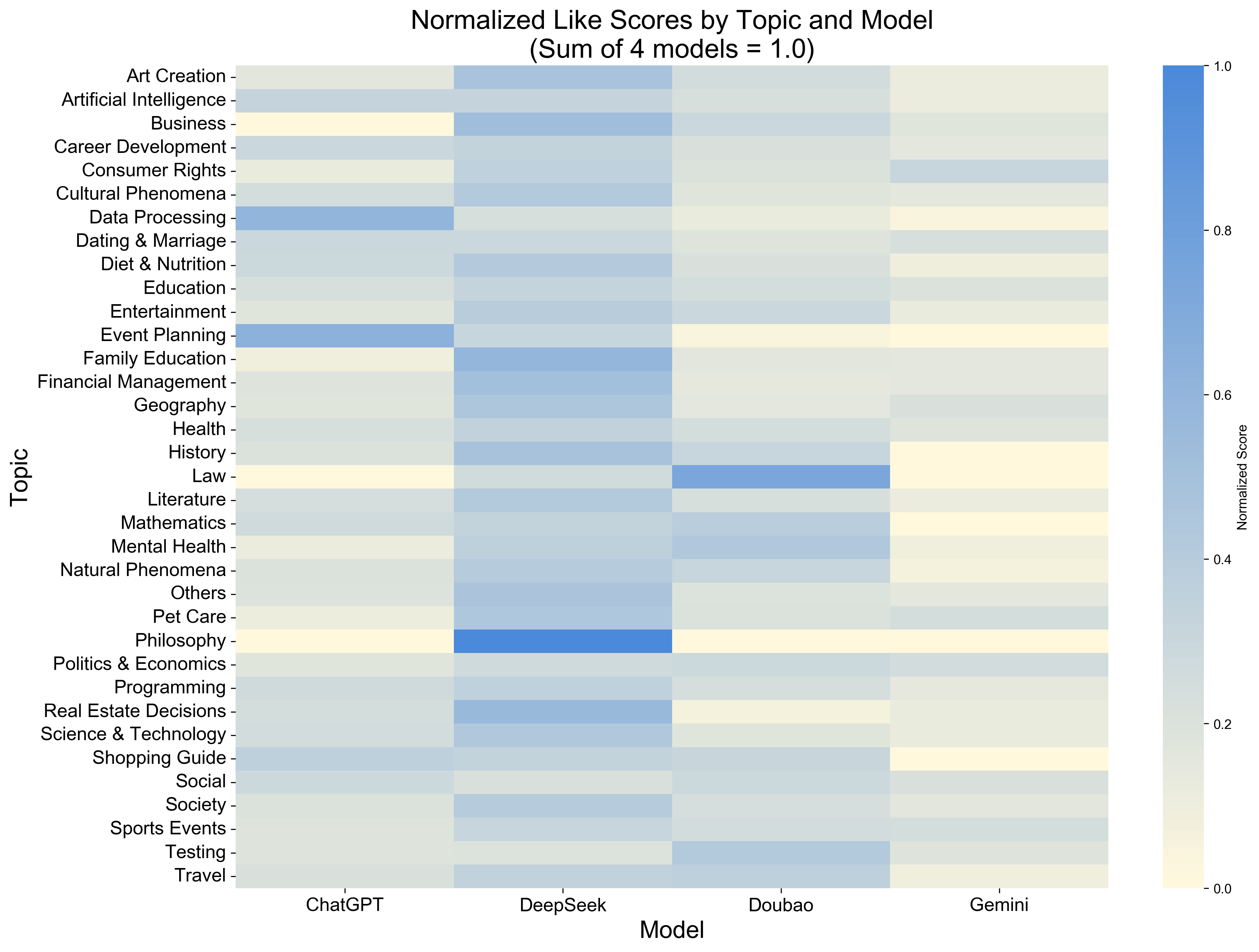} 
   
    \end{subfigure}
    \hfill
    \begin{subfigure}{0.8\textwidth}
        \centering
        \includegraphics[width=\textwidth, valign=t]{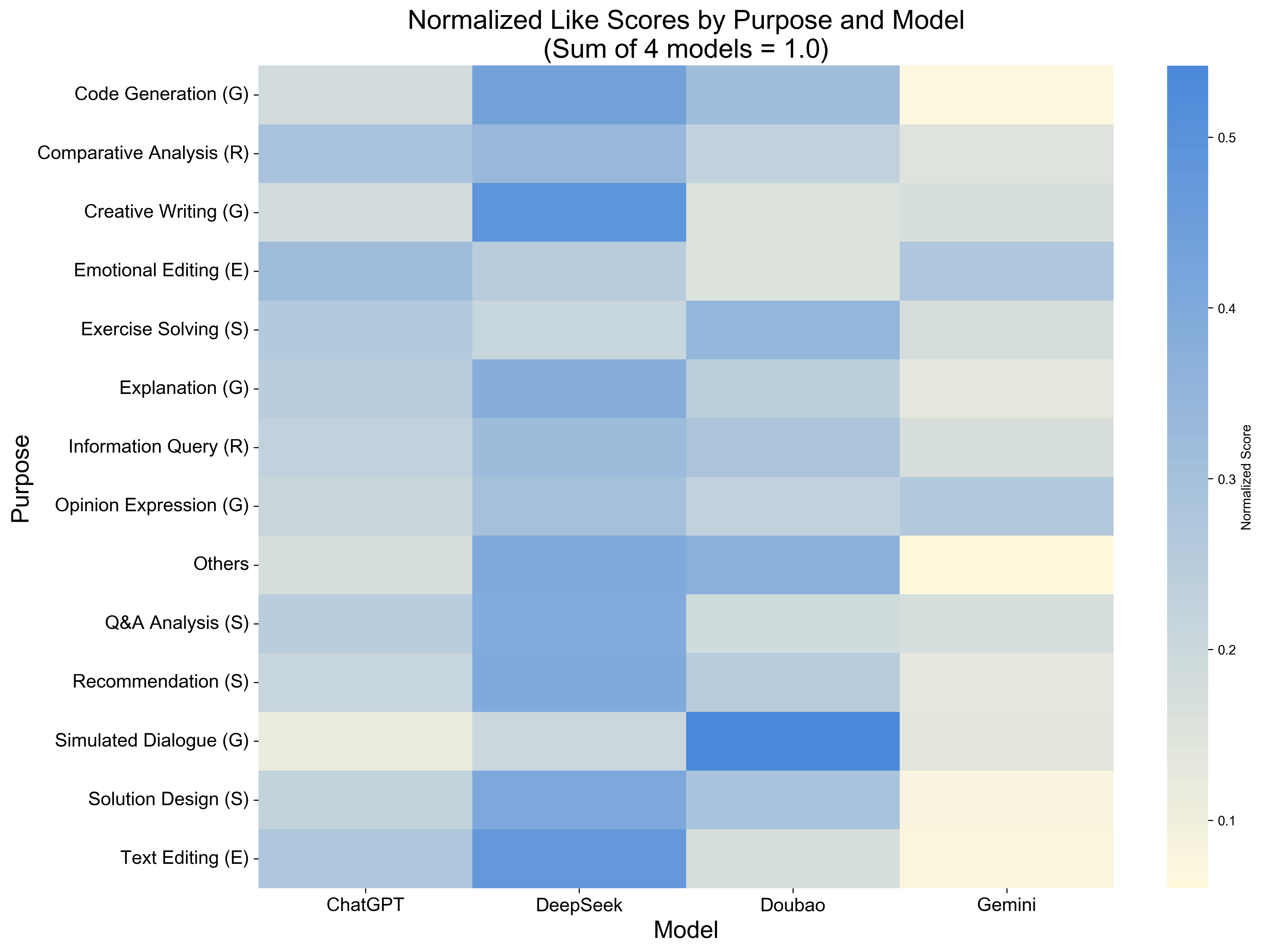} 
      
    \end{subfigure}
    \caption{Normalized model likelihood heatmap based on task type (Darker blue indicates higher score, Lighter yellow indicates lower score). A higher normalized score indicates a higher likelihood rate for the model on related task type.}
    \label{fig:task_like}
\end{figure}

In addition to task-based analysis, we also examined user habits regarding model usage. On average, users used 3.26 models per request (excluding regeneration), and 58.5\% of requests involved using all four models. Figure \ref{fig:heapmap} shows that each user's model preferences and usage frequency varied. We measured model performance using like and dislike rates, normalizing the scores within each task type (by dividing a single model's performance score by the sum of all performance scores for that task type). The result (Fig. \ref{fig:task_like}) shows that model performance indeed differed across task types. While Deepseek demonstrates superior performance in the domains of Philosophy and Family Education, ChatGPT holds an advantage in the areas of Event Planning and Data Processing. We also validated this finding through user feedback during the interviews.
% \yscomment{User-Model Invocation Heatmap, Task Topic Score Figure, Task Type Score Figure}

\subsection{Findings}
Based on our data analysis and user interviews, we derived the following conclusions:

\textbf{1. Information breadth, depth, and output differences drive users to try different models.}
Our data and user feedback indicate that model outputs vary in style and content across different usage scenarios, and user preferences and habits also differ. Some users found DeepSeek's responses detailed and structured, meeting their expectations, while others considered its answers overly verbose and preferred Gemini's concise style. Normal users typically lack prior knowledge about model characteristics and cannot determine which model is better suited for a specific problem or which output aligns more closely with their expectations before getting the results. Consequently, to obtain more comprehensive or suitable results, users tend to use multiple models simultaneously for comparison or attempt to integrate outputs from different models. This fundamentally addresses the user motivations for multi-LLM usage explored in RQ1. During interviews, we asked users to rate the necessity of using multiple models simultaneously on a 7-point Likert scale, yielding an average score of 5.53 (SD = 1.46).

\textbf{2. Task purpose influences the number of models users compare and their behavior.}
When asked about their expected number of models for comparison, 63.16\% of users stated that three to five models were sufficient for their needs, but also pointed out that this number is closely related to the task purpose. For tasks with relatively fixed or consistent answers (e.g., information retrieval questions), results from different models show little significant variation. Users only need one or two models to verify the results. However, for highly subjective tasks with diverse outputs, such as content generation, users prefer to examine results from multiple models to gain richer information. Furthermore, regarding the handling of model results, users expressed that if results differ significantly, they wish to expand all results simultaneously for quick browsing. Conversely, if model outputs contain overlapping or similar content, they prefer an intuitive way to identify the differences.

\textbf{3. The "Surprise" is a key factor enhancing user evaluation and acceptance.}
Combining user textual comments and interview feedback, we found that users particularly value "surprise" in model outputs—i.e., unexpected thorough considerations or unique information beyond their expectations. Users stated that if a model's output contains such "surprise," they are more inclined to adopt its results because "it makes me feel the model considered more comprehensively, and the output is more reliable." The pursuit of "surprise" also motivates users to proactively try multiple models and integrate results to obtain more comprehensive responses. Points 2 and 3 collectively address the influence of task and model characteristics on user behavior explored in RQ2.

\textbf{4. Users exhibit a "Browse-Compare-Integrate" behavior pattern.}
Users generally reported that after obtaining outputs from multiple models, they typically first browse the overall structure and then compare the similarities and differences between various outputs. When adopting results, besides directly selecting output from a specific model, many users actively integrate outputs from multiple models to form the final desired result. In existing commercial LLM interactions, they often need to switch between different pages or applications and finally integrate the results in an editing application, which significantly increases their operational and cognitive load. Although our current platform reduces the need for multi-page switching, users still expressed a strong demand for efficient comparison and integration features. This clearly identifies the multi-model interaction paradigms and existing pain points concerned in RQ3.

\subsection{Design Guideline}
\label{sec:guidelines}
Based on our experimental observations and user interviews, we derived three design guidelines aimed at enhancing the user experience and effectiveness of multi-LLM interaction systems:

\textbf{1. Provide context-aware intent recognition and model recommendation} 
The system should be capable of automatically identifying the task intent (e.g., content generation, content editing, problem-solving) and dynamically adjust the default number and type of models responding accordingly. For instance, models more proficient in the detected task type should be prioritized. Multiple models could be activated by default for creative tasks, while a single model might suffice for factual queries.

\textbf{2. Support differentiated parallel responses and flexible comparison schemes}  
The system should support parallel response generation from multiple models within a unified interface and allow users to easily expand, compare, and focus on different results. For tasks with significantly divergent outputs, parallel display should be provided to facilitate quick browsing. For tasks with similar results, the system should automatically summarize consensus points and highlight key differences to reduce users' cognitive load.

\textbf{3. Enhance result actionability and integrability} 
Effective tools and interfaces should be provided for users to act on model outputs, such as supporting segment selection and one-click merging of answers across models. Users should be able to flexibly combine outputs from different models to independently construct a final result that better meets their needs, thereby optimizing the "Browse-Compare-Integrate" behavior and reducing operational overhead.

\section{Platform Design}
Based on the observations and interviews from Experiment 1, we summarized three key design guidelines (Section \ref{sec:guidelines}). With these guidelines as our design objectives, we aimed to develop a next-generation multi-LLM interaction platform that not only enhances interaction efficiency but also effectively reduces users' cognitive load. Building upon the prototype from Experiment 1, we introduced the following three new features:
\begin{itemize}
    \item 1. Task Type Recognition and Model Recommendation: Automatically identifies the type of user query and recommends suitable models.
    \item 2. Enhanced Model Output Presentation Mechanism: Visually presents differences and consensus among models based on their generated content.
    \item 3. Flexible Multi-Model Result Integration: Provides both automatic and manual result integration functionalities.

\end{itemize}

\subsection{Workflow}
\begin{figure}[htbp]
    \centering
    \includegraphics[width=0.9\linewidth]{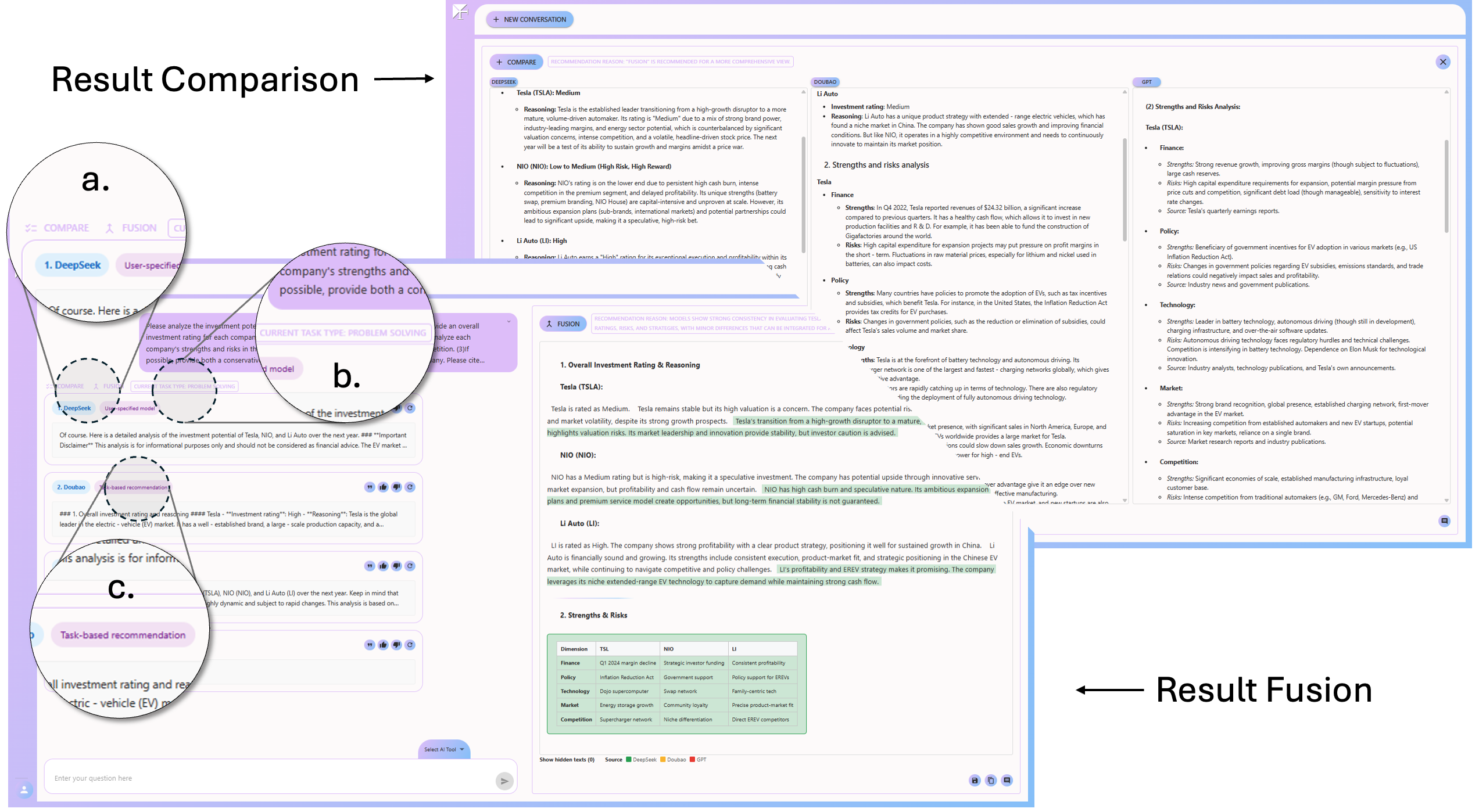}
    \caption{LLMartini provides two comparison modes. The top image shows the comparison results, and the bottom left image shows the fusion results. Zoomed view highlighting (a) fusion/comparison buttons, (b) current task type (e.g., “Problem-Solving”), and (c) reason for tool recommendations}
    \label{fig:workflow}
\end{figure}

After a user submits a query, the system backend first performs real-time intent recognition and task type classification. This type of label is displayed in the interaction interface (Fig. \ref{fig:workflow} b.) in real-time to enhance the explainability of the system's behavior and users' perceptual transparency. Based on the identified task type, the system dynamically adjusts its model dispatching strategy. This includes recommending the order for invoking models and automatically executing the corresponding number of models according to the task purpose. Recommendation reasons (Fig. \ref{fig:workflow} c.) are attached to the summary cards generated by each model. In addition to the models invoked automatically by the system, users can also manually trigger other models to run.

After obtaining the model results, the system provides an "Fusion" function (Fig. \ref{fig:workflow} a.). Upon clicking the integration button, users can select the models whose outputs they wish to integrate in a pop-up interface. The system will then synthesize the user's selected results, the original query, and the task type to algorithmically recommend an appropriate presentation method, specifically divided into two modes:

\begin{itemize}
    \item Result Comparison: Suitable for situations where outputs differ significantly. It presents the detailed results of multiple models in a side-by-side, synchronized scrolling format, facilitating intuitive comparison.
    \item Result Fusion: Suitable for results with overlapping content or similar semantics. It processes and integrates the outputs, explaining at the top of the interface for recommending this mode.
\end{itemize}
Furthermore, we retained a direct button for switching to the result comparison view, allowing users to easily access the comparison page if desired.

% \yscomment{llm card, comparison, fusion}

\subsection{Implementation}
\subsubsection{Task Objective Awareness and Dynamic Model Dispatching}
Results from Study 1 indicate that the task category significantly influences users' behavioral patterns when interacting with multiple models. To address this, we introduced a mechanism for dynamically dispatching models based on the task objective. When a user submits a request, the system utilizes the task classifier constructed in Experiment 1 to identify the task objective (including major and minor categories) in real time and records it. The model dispatching strategy is primarily formulated based on the performance ranking of models under each major task category, which is calculated from the aggregated user data of Study 1. We chose to make recommendations at the major category level (rather than minor categories) because some minor categories have limited sample sizes and uneven user distribution, which could introduce user bias and affect the generalizability of recommendations.

\begin{figure}[htbp]
    \centering
    \includegraphics[width=0.9\linewidth]{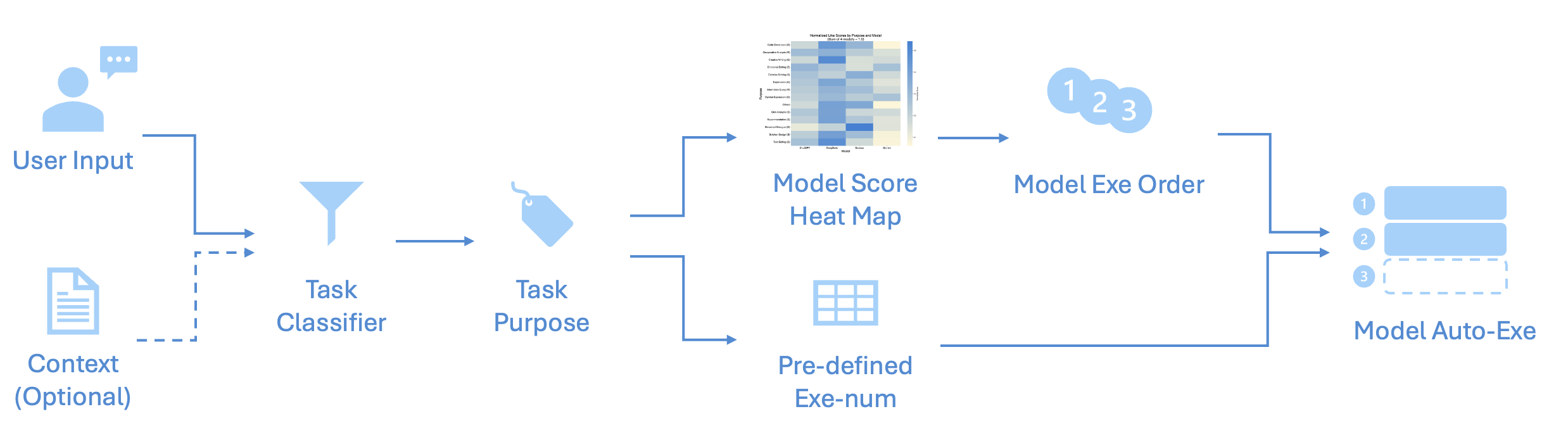}
    \caption{LLMartini Dynamic Model Dispatching. First, the task purpose is analyzed based on user input and the current context. Then, the number and order of model executions are matched according to the task purpose. Finally, the model is automatically executed.}
    \label{fig:dispatching}
\end{figure}
During the dispatching process, models actively specified by the user still receive the highest priority and are placed at the top of the queue. The system predefines the number of models to be automatically executed for each major task category: the "Content Generation" category, which demands the highest creativity, automatically invokes all four models, while the other major categories default to executing the top two best-performing models.

\subsubsection{Adaptive Integration and Comparative Display}
After the user selects the model results to be integrated, the system initiates an adaptive integration process by synthesizing the user's request, task objective, and model outputs. The integration process is divided into two stages: preliminary integration and integration enhancement.
\begin{figure}[htbp]
    \centering
    \includegraphics[width=0.9\linewidth]{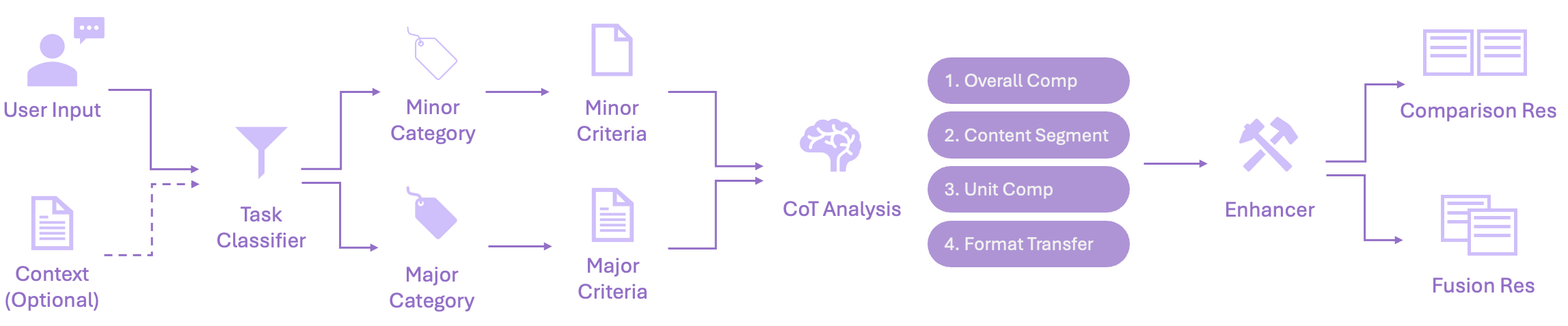}
    \caption{LLMartini Fusion and Comparison Pipeline. First, the task purpose is analyzed based on user input and the current context. Then, the corresponding evaluation criteria are obtained based on the task purpose. A CoT approach decomposes the model output into comparison units, which are then integrated under the guidance of the evaluation criteria to produce the final result.}
    \label{fig:fusion}
\end{figure}

We employ a Chain-of-Thought prompting approach to achieve preliminary integration. Since users have different focuses for different tasks, the system maintains a two-tiered set of integration criteria: major category integration criteria and minor category integration criteria. The major category integration criteria define the core focus points for each type of task. For example, the "Problem Solving" category emphasizes solutions and logic, while the "Content Editing" category focuses on modification quality and style consistency. The minor category integration criteria are refined to specific subcategories, providing standards for determining differences between results and their priorities.

In the preliminary integration, the integrator first evaluates the overall differences between results based on the major category criteria. If the differences are too significant, it directly recommends a comparative display; otherwise, it attempts integration. For model outputs requiring integration, the system first structures the outputs into blocks according to the task type. For instance, in information retrieval tasks, model outputs can be divided into modules such as "Task Comprehension," "Information Provision," "Summary," and "Resource References." The system allows for differences in module completeness across models, as varying levels of completeness are an important reflection of differences between models. Content within each module is further segmented semantically into units that can be compared or integrated.

Subsequently, the integrator judges whether semantic units generated by different models are similar based on the minor category criteria. Units originating from a single model are retained; units generated by multiple models are either merged or retained in parallel based on semantic similarity. Finally, the system converts the output into a structured JSON format, where each unit corresponds to a list. The list elements annotate the source model and the corresponding content.

To implement this process, we developed dedicated prompt templates for each major task category, embedding the integration criteria and segmentation logic for that category. We also introduced preferences for integration versus comparison (e.g., content generation tasks prefer comparison, while content editing tasks lean towards integration). Additionally, we included relevant examples to help the integrator understand how to segment and differentiate model differences. Minor category criteria are dynamically concatenated into the major category prompt during integration to provide a more refined judgment basis.

After preliminary integration, direct concatenation of some units may lead to a disjointed expression. Therefore, we introduce an enhancer to optimize and validate the results. Firstly, it improves textual coherence and readability, enhancing the quality of the integrated output. Secondly, it performs secondary validation to ensure the rationality of each unit segmentation, splitting or merging units as necessary to improve display effectiveness and user experience.

\subsubsection{Multi-LLM Result Integration Interface}
In the integration display view, the system divides content generated by different models into paragraph or sentence-level fragments based on semantic units. The source model of each fragment is distinguished by color highlighting, with a legend at the bottom indicating the color-model correspondence. If a segment of content is generated by only a single model, it is highlighted in the corresponding color. If multiple models generate semantically similar content, it is automatically merged and not highlighted to reduce visual clutter and indicate high consensus and reliability. If different models generate relevant but semantically distinct content, both versions are retained, and a switching mechanism is provided—users can browse different expressions of the same content using left and right arrows that appear upon hovering.

\begin{figure}[htbp]
    \centering
    \includegraphics[width=0.9\linewidth]{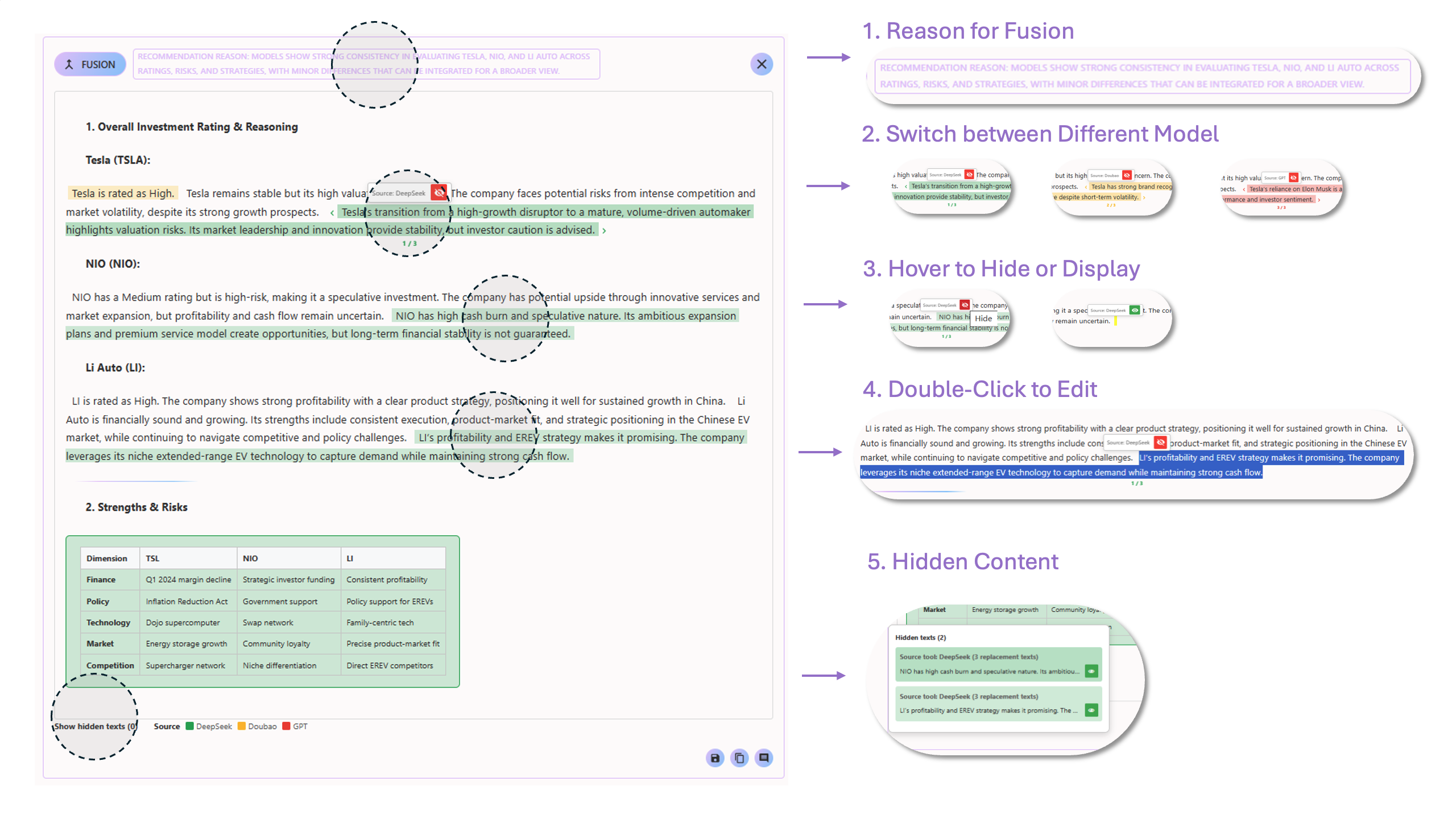}
    \caption{LLMartini Fusion Interface. We provide users with rich fusion interactions. Users can view fusion reasons (1), switch results from different model sources (2), hide or show models (3), edit results (4), and uniformly view hidden content (5).}
    \label{fig:fusion_inter}
\end{figure}

Additionally, hovering displays the model name and allows users to hide the current paragraph. Hidden content is marked with a yellow placeholder, and users can restore it at any time via hover actions or the "Hidden Content" panel in the lower left corner of the page. Users can also double-click any paragraph to enter edit mode directly, allowing free adjustment of the content. Through these features, users can quickly identify differences and unique contributions from various model outputs, conveniently switch, hide, or edit content, and efficiently construct an output that meets their needs. When the user clicks the "Copy" button, the system automatically integrates all currently visible and edited content in the interface and writes it to the clipboard for subsequent use.

\section{Study2: System Usability and User Evaluation}
To evaluate the effectiveness, usability, and user satisfaction of \projectName{} in real-world usage scenarios, we conducted a user experiment to assess system usability and gather user feedback. We aimed to address how \projectName{} performs compared to traditional multi-tab or multi-window interaction methods in terms of task efficiency, user experience, and cognitive load.

\subsection{Study Design}
\subsubsection{Participants}
We recruited 18 participants (8 male, 10 female, mean age = 24.8, SD = 3.3). All participants were selected from a sample pool distinct from Study 1 to ensure evaluation independence. To cover as much task diversity as possible, we selected participants with diverse backgrounds and varying levels of LLM experience.

\subsubsection{Tasks and Procedure}
We defined the method where users manually switch between tabs or applications to interact with different models and manually combine the outputs as the baseline. The \projectName{} platform served as the experimental condition.

The experiment employed a within-subject design; each user experienced both interaction methods. To counterbalance order effects, half of the users started with the baseline method, while the other half began with \projectName{}.

Each participant was required to complete a series of predefined tasks. The experiment consisted of two sessions, each using one interaction method. Within each session, users were asked to complete three distinct tasks with different topics under each of the four different task purposes (Content Generation, Content Editing, Information Retrieval, Problem Solving), resulting in 12 tasks per session. Regarding task topics, those obtained from Study 1 were divided into three sets based on their frequency of occurrence: high(more than 25 times), medium(10-25 times), and low(less than 10 times). The topic for each distinct task was randomly selected from one of these sets according to the user's own background, ensuring minimal topic similarity between users while guaranteeing coverage of all topics. Users needed to conceptualize the specific task based on the given objective and topic and complete it through interaction. In total, each user completed 2 methods × 4 task objectives × 3 task topics = 24 interactive tasks.

At the beginning of the experiment, the facilitator introduced the background and task content. Before using \projectName{}, the facilitator demonstrated its main features, including multi-model response generation, result comparison, and result integration, using a simple query. Users then proceeded through the two experimental sessions. Task completion time was recorded by the facilitator, excluding the time spent conceptualizing the question and waiting time for model output. There is 10 10-minute break between sessions. After completing all the tasks, users filled out the NASA-TLX \cite{hart1988nasatlx} questionnaire and the system SUS \cite{brooke1996sus} questionnaire for both methods. This was followed by a semi-structured interview, inviting users to share their experiences and feedback. The entire experiment typically lasted from 90 to 120 minutes. Participants were compensated with a \$20-30 gift card upon completion.

\subsection{ Results and Analysis}
\subsubsection{Quantitative Analysis Results}
We calculated usability and cognitive load metrics for both methods. The average task completion times for LLMartini and the baseline were 2.59 minutes (SD = 1.80) and 3.40 minutes (SD = 1.48), respectively. Our system significantly reduced the time required to complete tasks ($p = 0.015 < 0.05$). This indicates that LLMartini effectively enhances task efficiency by reducing context switching and providing an integrated workflow.

The SUS scores for LLMartini and the baseline were 75.56 (SD = 11.32) and 49.72 (SD = 11.84) (out of 100), respectively. According to the SUS rating scale, \projectName{}'s score falls within the Good range and is significantly higher than that of the baseline method ($p < 0.05$). LLMartini shows no significant difference from the baseline method in terms of Learn Quickly and Need to Learn. This is because the baseline method is generated from users' daily usage habits, resulting in a lower learning cost. The absence of a significant difference between the two methods also indicates that LLMartini itself has a very low learning cost. In contrast, LLMartini scores significantly higher than the baseline in Need Support. This is because the baseline method requires almost no guidance, whereas LLMartini introduces a new fusion interface, and users who are new to the platform may require some functional introductions to help them quickly understand its capabilities. Additionally, LLMartini significantly outperforms the baseline method across all individual criteria of usability.
\begin{figure}[htbp]
    \centering
    \includegraphics[width=0.9\linewidth]{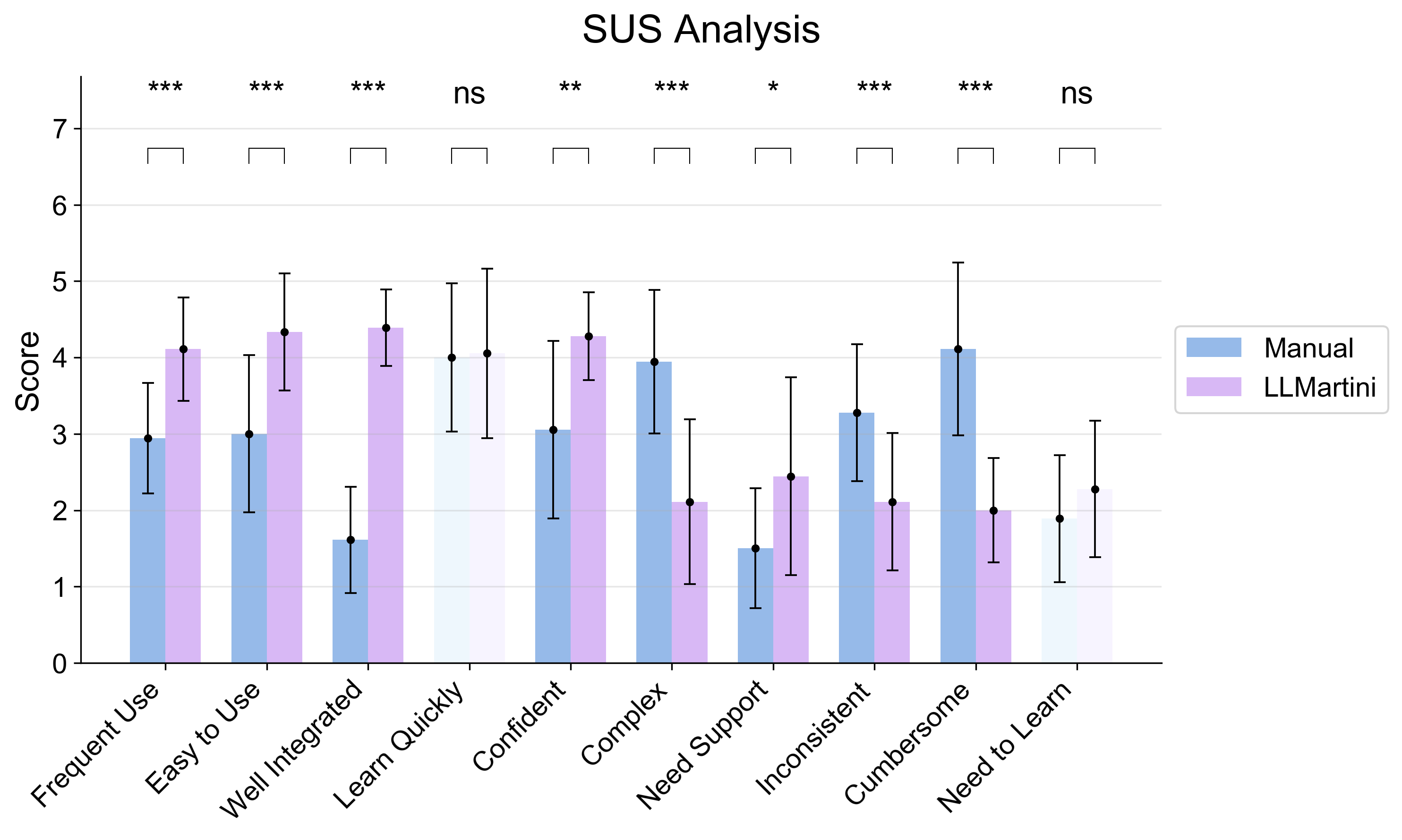}
    \caption{SUS scores and significance analysis results of LLMartini and baseline.}
    \label{fig:sus}
\end{figure}

Regarding cognitive load, the NASA-TLX scores for \projectName{} and the baseline were 40.58 (SD = 14.65) and 59.66 (SD = 17.13) (out of 100), respectively. LLMartini significantly reduced users' cognitive load ($p < 0.05$). Users reported that "not having to switch between apps" and visualization of differences greatly eased their comparison burden. For the Performance score, there was no significant difference between the baseline method and LLMartini. This is because the task completion criterion was defined as users having obtained a model fusion result they found acceptable for their task. Consequently, users were able to achieve their desired output with both methods, leading to comparable scores in the performance category. In contrast, LLMartini demonstrated a significant reduction in burden across all other individual burden metrics.
\begin{figure}[htbp]
    \centering
    \includegraphics[width=0.8\linewidth]{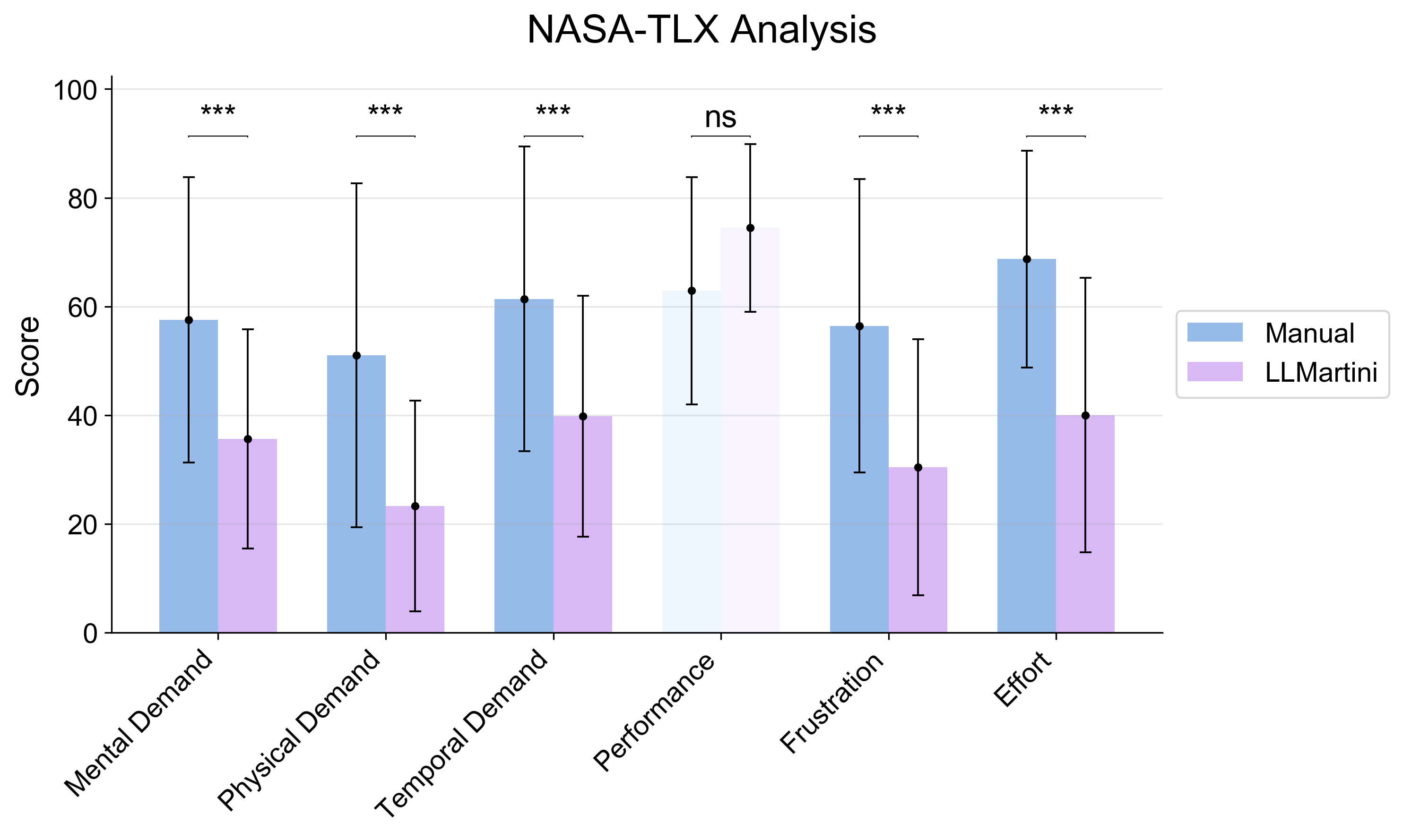}
    \caption{NASA-TXL scores and significance analysis results of LLMartini and baseline}
    \label{fig:nasa}
\end{figure}

\subsubsection{User Preferences and Behavioral Patterns}
During the interviews, we recorded users' usability evaluations and willingness to use both systems on a 7-point Likert scale. LLMartini scored significantly higher than the baseline in both usability (5.89 vs 2.67) and willingness to use (6.17 vs 4.06) ($p < 0.05$).
\begin{figure}[htbp]
    \centering
    \includegraphics[width=0.8\linewidth]{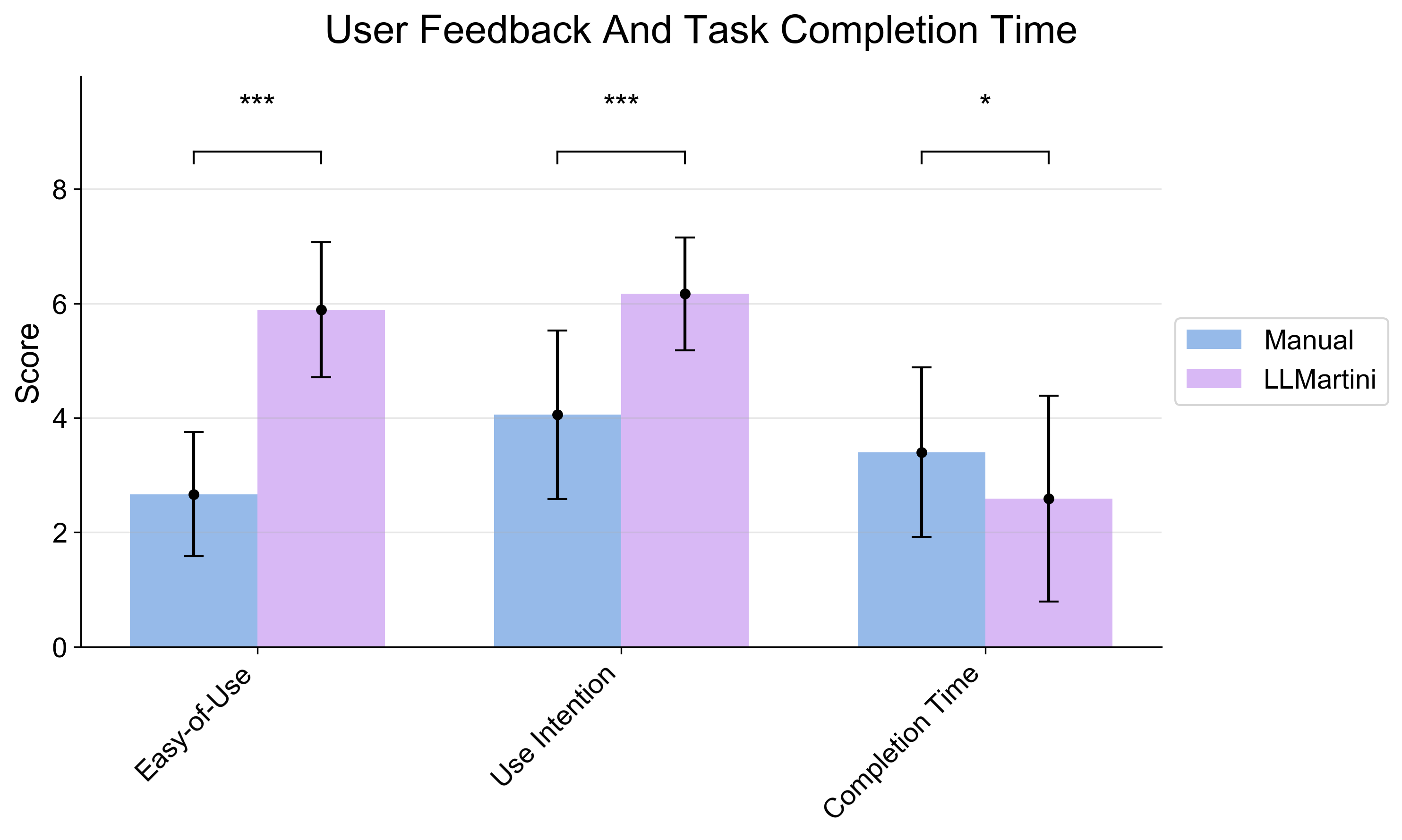}
    \caption{User feedback scores and Completion time significance analysis results of LLMartini and baseline.}
    \label{fig:use}
\end{figure}

All participants agreed that LLMartini’s approach of interacting with multiple models simultaneously and automatically integrating their outputs was an exciting design. Participant 14 further commented that the platform offered more than just combining results from four models: “Through functions such as quoting, simultaneous invocation, and fusion, I can achieve collaboration among multiple language models—for example, using outputs from two models as context for the next round, then fuse the output from the next round.” This made users feel that LLMartini provides greater potential for multi-model collaboration.

We observed that different users focused on different aspects when using LLMartini. Users with less experience in LLM interaction tended to appreciate the speed and convenience of the fusion function. Participant 9 mentioned that LLMartini’s automatic fusion reduced their decision-making burden, eliminating the need to manually compare and select model results, while quickly delivering more comprehensive outcomes. On the other hand, users who frequently use LLMs valued the flexibility offered by the system, which enhanced their efficiency when combining model outputs.

However, some users noted that when fusing lengthy model outputs, LLMartini might omit certain details. This loss of information could occasionally increase their cognitive load, sometimes even requiring them to manually review the original results. They expressed a desire for the ability to configure parameters of the fusion function—such as increasing the weight of a specific model or adjusting the granularity of fusion.

\subsubsection{Unexpected Usage Behaviors and Design Implications}
In addition to the user usage we expect, we observed several interesting, unexpected usage patterns:
\begin{itemize}
    \item [1.] "Exploratory Comparison" Behavior: Some users intentionally selected models with the most divergent output styles for comparison, not to solve a specific problem but to "understand the performance differences between different models" or "test the boundary of comparison and fusion".

    \item [2.] Novel Use of Integration Feature: Some users utilized the integration feature as a "quality filter," judging the reliability of information by observing which content was commonly generated by multiple models (the automatically merged sections).
    
    \item [3.] Session Knowledge Accumulation: Some users employed the fusion results as contextual information in subsequent interactions. They viewed LLMartini as a platform for simultaneous collaboration with multiple LLMs, where each interaction generated collective knowledge through fusion and propagated it to all models.

\end{itemize}
Some expert users suggested the system should record historical comparison results to form a personalized "model characteristic knowledge base." They expected the system to record their interaction habits and learn how to achieve better fusion from their interactions.
These findings offer new insights for multi-LLM interaction:
a) There is a need to support more flexible comparison and fusion strategies; b) A more adaptable contextual reference mechanism should be provided; c) The introduction of long-term learning and personal adaptation capabilities should be considered.

% These findings provide important implications for future design: a) support for more flexible comparison strategies is needed; b) a credibility visualization mechanism should be provided; c) long-term learning and personal adaptation features should be considered.

The results of Study 2 confirm the effectiveness of LLMartini in enhancing the efficiency of multi-model interactions. Compared to the traditional method, our system performed better not only in objective metrics (task time, error rate) but also showed significant improvement in subjective user experience (satisfaction, cognitive load).

\section{Discussion}
\subsection{From Tool to Collaborator}
The initial inspiration for designing LLMartini came from users' need to utilize multiple models simultaneously. However, as our research progressed, we gradually discovered that LLMartini is no longer just a tool for improving efficiency in user interactions with multiple models. 

When users interact with multiple models, beyond comparing and integrating results, there are more complex collaboration patterns at play. Certain complex tasks require users to select different models at different stages to achieve better outcomes. In such scenarios, the user's interaction resembles working with a team of individuals skilled in different areas. This "team collaboration" style of interaction prompted us to rethink LLMartini’s system positioning—it is no longer merely a platform for multi-model invocation and integration but rather a mediator of cognitive collaboration. Through LLMartini, users gain insights into the capabilities of different models, dynamically assign tasks based on objectives, and continuously adjust collaboration and information sharing among models during execution. For example, in a creative generation task, a user might first use one model for brainstorming, then another for logical validation, and finally call upon a third model to optimize the expression style. The entire process is no longer a simple parallel invocation or result integration but a highly dynamic, temporally structured intelligent collaborative behavior.

This discovery also hints at new possibilities for future AI systems in supporting complex tasks: systems should not only focus on improving model performance but also pay attention to how to help users flexibly organize multiple model capabilities, making them a natural extension of the user's thought process. The paradigm of user interaction with multiple models should shift from "using tools" to "leading a team," thereby unleashing the collaborative potential of human and model collectives in more complex intelligent tasks.

\subsection{More Powerful Model vs Human-Centered System}
While attempting to help users improve the efficiency of integrating model outputs, a question has consistently lingered around us: Could a single, more powerful model change this? A phenomenon observed in the first user study helped us answer this question: two users formed opposite judgments about the same model's capabilities. This revealed the significant variability that user preferences and personalization bring to the interaction experience. Although AI capabilities continue to advance across various aspects, it is challenging for any single model to meet the needs of every user.

The "power" of a single model does not equate to comprehensive coverage of user needs. Users' expectations and evaluation criteria for models are often highly subjective, stemming from their task contexts, cognitive habits, and even aesthetic preferences. For example, one user might prioritize the interpretability of generated results, while another may value creativity or execution efficiency more. Such differences mean that even when facing the same model, different users can have vastly different experiences and levels of satisfaction.

This further validates the fundamental significance of human-centered interaction systems like LLMartini: their goal is not merely to pursue the "optimal" performance of AI but to provide support from the user's perspective, enabling users to autonomously construct results that best suit their current tasks and personal preferences. During user interviews, we also received suggestions about a personalized fusion method. Therefore, future interaction designs should place greater emphasis on incorporating personalized factors into the system architecture, such as introducing more fine-grained user preference perception mechanisms and providing adaptive guidance for model recommendation and combination strategies. Only in this way can a deeper alignment between the diverse needs of different users and the technical capabilities of models be achieved.

\subsection{Modalities beyond Text}
Current large models have progressively achieved multimodal and even Agent capabilities. Although LLMartini has achieved effective integration and comparison results in the textual modality, it has also prompted us to delve deeper into the integration methods for other modalities. Compared to the structured and stable nature of text, the fusion mechanisms for modalities such as images, audio, and even video are far more complex.

From the perspective of output form, text inherently possesses a strong structural quality, allowing semantic units to be segmented and reorganized through methods like sentence or paragraph division, typically without compromising the overall meaning. In contrast, content such as images and audio often exists within a continuous, high-dimensional, and implicitly semantic structure. Simple cutting or fragmentation can easily disrupt their informational integrity and aesthetic consistency. For example, separating regions of an image may result in the loss of global compositional intent, while segmenting audio may break its emotional coherence.

Furthermore, from the perspective of user interaction, the editing and integration of multimodal content present entirely new challenges. Users not only need to address alignment issues between different modalities (such as image-text associations or audio-visual synchronization) but also must manage the manipulation and organization of multiple media types simultaneously within the interface. Traditional text-centric interaction models are difficult to directly apply to multimodal contexts. How to design an intuitive and efficient multimodal workflow—enabling users to flexibly compare, combine, and even create cross-modal content—has become a critical issue requiring urgent exploration.

Therefore, future multi-model interaction platforms like LLMartini must not only architecturally support the input and output of heterogeneous modalities but also rethink the design of the entire interaction paradigm. Potential directions include developing interfaces that support multimodal annotation and dynamic preview, introducing non-destructive editing operations (such as layering, masking, and timeline editing), and even providing visual representations of cross-modal semantic associations. These advancements would help users effectively exercise creative judgment and maintain control in significantly more complex environments.

\subsection{Limitation and Future Work}
While LLMartini presents considerable advantages for facilitating multi-LLM interactions, our current research is subject to several limitations that suggest productive avenues for further investigation.

First, the existing framework for task classification and model recommendation is derived from a relatively small and heterogeneous sample (Study 1, N=19), which may introduce selection bias and constrain representativeness. Due to substantial variation in participant backgrounds, several task categories were underpinned by limited responses, potentially allowing the preferences of a narrow user subset to disproportionately sway model recommendations. A more robust and personalized approach should integrate a triad of factors: task characteristics, objectively evaluated model capabilities, and individualized user preferences. Future efforts will focus on expanding the dataset to improve the validity and generalizability of the classification and recommendation mechanisms.

Second, the evaluation of the system was conducted under controlled laboratory conditions (N=18). Although the findings yielded encouraging outcomes, they necessitate validation through larger-scale longitudinal studies conducted in authentic settings. Such research is critical to capturing emergent patterns of adoption, adaptive use over time, and the evolving needs of users when interacting with multi-LLM systems in real-world contexts.

Finally, the present study concentrates primarily on core functionalities such as output comparison and composition. However, participant feedback throughout the study underscored a compelling demand for more sophisticated interfaces that support advanced collaborative workflows among multiple LLMs. Moreover, extending this work to incorporate multimodal outputs (e.g., text, image, audio) and agent-based interactive frameworks represents a promising yet underexplored territory with substantial theoretical and practical implications.

\section{Conclusion}
In conclusion, to address the inefficiencies of manually comparing and combining outputs from multiple LLMs, we introduced LLMartini, a novel interactive system that enables seamless comparison and user-controlled composition through semantic segmentation and visual highlighting. A user study demonstrated that LLMartini significantly outperforms conventional methods by reducing task time, cognitive load, and increasing satisfaction. Our work highlights the importance of human-centered design in leveraging the complementary strengths of diverse AI models, providing a foundation for more efficient and creative multi-LLM interactions.

\begin{acks}
To Robert, for the bagels and for explaining CMYK and color spaces.
\end{acks}

\bibliographystyle{ACM-Reference-Format}
\bibliography{LLMartini}

\appendix

\end{document}